\documentclass[12pt]{article}

\usepackage{newtxtext,newtxmath}

\usepackage{graphicx}
\usepackage{float,subcaption,siunitx}
\usepackage{xcolor}
\usepackage{bm}
\usepackage[letterpaper,margin=1in]{geometry}

\renewenvironment{abstract}
{\quotation}
{\endquotation}

\date{}

\makeatletter
\renewcommand{\fnum@figure}{\textbf{Figure \thefigure}}
\renewcommand{\fnum@table}{\textbf{Table \thetable}}
\makeatother

\usepackage{scicite}

\usepackage{amsmath}

\usepackage{url}

\newcommand{\SRO}		{Sr${}_2$RuO${}_4$}
\newcommand{\UPt}		{UPt${}_3$ }
\newcommand{\Hethree}		{$^{3}${He} }

\def\scititle{
Visualization of vortex sheets and half quantum vortices in the chiral odd-parity superconductor ${\mathrm{UPt}}_{3}$
}
\title{\bfseries \boldmath \scititle}

\author
{P. Garc\'ia-Campos$^{1}$, 
V.O. Dolocan$^{1,3}$,
M.-K. Arfaoui$^{1,2}$,
D. Aoki,$^{2,5}$, 
A. D. Huxley$^{4}$, \\
K. Hasselbach$^{1\ast}$\\
\\
	\normalsize{$^{1}$Universit\'{e} Grenoble Alpes, CNRS, \& Institut N\'{e}el,  38042 Grenoble, France}\\
	\normalsize{$^{2}$Universit\'{e} Grenoble Alpes, CEA, IRIG, PHELIQS, 38000, Grenoble, France
	}\\
 \normalsize{$^{3}$Aix Marseille Univ, Université de Toulon, CNRS, IM2NP, Marseille, France}\\
	\normalsize{$^{4}$School of Physics and Astronomy and Centre for Science at Extreme Conditions,}\\
 \normalsize{The University of Edinburgh, Edinburgh EH9 3FD, United Kingdom}\\
    \normalsize{$^{5}$Institute for Materials Research, Tohoku University, Ibaraki 311-1313, Japan}\\
	\normalsize{$^\ast$To whom correspondence should be addressed; E-mail: klaus.hasselbach@neel.cnrs.fr}
}

\date{}

\begin{document}

\baselineskip24pt


\maketitle 

\begin{abstract} \bfseries \boldmath

	Superconductivity is characterized by vanishing electrical resistance and magnetic flux expulsion. 
 For conventional type II superconductors, the magnetic flux expulsion is incomplete in an applied magnetic field above a critical value and magnetic flux penetrates the bulk of the superconductor in discrete quantized magnetic flux tubes (\textit{vortices}), each carrying a single quantum of flux ($h/2e$). 
 Investigating the unconventional superconductor UPt$_{3}$ with a scanning superconducting quantum interference device (SQUID), we observed mobile half-quantum vortices together with one quantum vortices. Cooling the material under a higher magnetic field revealed the presence of lines of magnetic contrast resembling domain walls. These observations agree with theoretical predictions for chiral superconductivity with a two dimensional complex order-parameter with sheets of half-quantum vortices separating domains of opposite order-parameter chirality.

\end{abstract}

\maketitle
\section*{Introduction}

A superconducting state is a coherent collective state of a macroscopic number of electrons, described by a multielectron complex wavefunction. The wavefunction is constructed from bound electron-electron pairs called Cooper pairs that  
are anti-symmetric under exchange of their constituent electrons. If the crystal possesses inversion symmetry the pair wavefunction factorizes as a product of spin and spatial parts. In conventional superconductors, 
the spatial part has the full symmetry of the crystal point group and the pair is in a spin singlet state ($S$=0) \cite{Mineev_introduction_1999,Poole2007}. 
In unconventional superconductors the spatial part of the Cooper pair wavefunction is no longer invariant under the crystal point group. 
If the parity of the spatial part is even the superconductivity has $S=0$, while an odd parity requires spin $S=1$. 
For some point groups unconventional superconductors with multi-component order parameters are possible.  These states open up new possibilities, including the formation of fractional vortices and domains\cite{Volovik1976}.


Unconventional pairing was first discovered in superfluid $^{3}${He}: of particular relevance to our study of UPt$_3$ is the A-phase of $^{3}${He}, which occurs just below the normal-superfluid transition temperature 2~mK under pressure $P>20 \text{bar}$. This has a chiral $p$-wave ($p_x \pm ip_y$) order parameter with $S=1$ that breaks time-reversal symmetry and has nontrivial topology\cite{Volovik2003}. 
UPt$_3$ is an example of a heavy-fermion metal in which the conduction electrons are strongly correlated by spin fluctuations leading to a greatly increased electronic specific heat. Although many heavy-fermion metals are unconventional superconductors  \cite{Joynt2002}, UPt$_{3}$ stands out in having three distinct superconducting phases \cite{Hasselbach1989,Adenwalla1990,Bruls1990,Dijk1993} at ambient pressure in moderate magnetic fields. 
In zero applied magnetic field, cooling from the normal state, the A-state appears at T$_c^+$ ($T_c^+$ is sample dependent in the range 510-560 mK for samples showing sharp transitions). This is followed by a transition to the B-phase at T$_c^-$ where $T_c^+-T_c^-=50-80\text{ mK}$.
The C phase appears in magnetic fields $H \parallel c >0.4 ~\text{T}$ and is not accessed in our measurements. The multiple superconducting phases point to a multi-component order parameter with the separation between $A$ and $B$ states attributed to a weak local deviation from hexagonal crystal symmetry;  only a single component of the order is then present in zero field in the A-phase. Evidence that the B-phase breaks time reversal symmetry (TRS) comes from (i) small-angle neutron scattering (SANS) from the vortex lattice following different field paths \cite{Avers2020}, (ii) internal fields detected with muon spin rotation ($\mu$SR) on poorer quality samples \cite{Luke1993} (the fields detected with $\mu$SR may be induced by crystalline defects since they are not found in better quality samples \cite{Reotier1995}) (iii) a Polar Kerr effect\cite{Schemm2014} and (iv) Josephson interferometry measurements\cite{Strand2009}.  The low temperature B-phase is then a TRS breaking chiral state analogous with $^{3}${He}-A. Such states are predicted to host novel chiral textures including domain walls decorated by half-quantum vortices (vortex sheets) separating regions of Cooper pairs with opposite chirality \cite{Volovik1985,Sigrist1989,Sigrist1999,Volovik2003,Matsunaga2004a,Ichioka2005,Kallin2016}. For UPt$_3$ a chiral TRS breaking B-phase is possible for different order parameter symmetries (irreducible representations) having different nodes in the gap function. The choice of order parameter has to explain the observed field-temperature phase diagram, and the power law dependence of various physical quantities\cite{Joynt2002,Machida2012,Gannon2015} as well as more direct Josephson interferometry measurements. 
A predominantly $f$-wave, odd-parity order parameter belonging to the $E_{2u}$ irreducible representation of the $D_{6h}$ point group\cite{Sauls1994,Nomoto2016,Yanase2016,Yanase_mobius_2017} appears to offer the most comprehensive explanation of these measurements.
Experimentally, vortex sheets between anti-phase domains were detected in bulk $^{3}$He A-phase by nuclear magnetic resonance (NMR) spectroscopy in a rotating cryostat \cite{Parts1994}. Half-quantum vortices have also been observed, but only in a nanoconfined polar phase of $^{3}$He \cite{Autti2016}. Half quantum vortices in a similarly confined distorted A-phase have not yet been seen experimentally, but are predicted to carry Majorana zero modes (a zero energy state bound to vortex cores that is an equal superposition of a conventional fermion and hole). Vortices with Majorana zero modes obey non-Abelian statistics and could in principle be braided to make a topological quantum computer\cite{Autti2016}. The theoretical determination of whether half quantum vortices on domain walls in UPt$_3$ carry Majorana zero modes is not yet clear and requires knowledge of the actual order parameter (and the known Fermi surface). If half quantum vortices in UPt$_3$ are found to have Majorana zero modes they would be much easier to manipulate than those in superfluid $^3$He.


Besides $^3$He, half-quantum vortices have been observed in engineered geometries of spin-singlet high T$_c$ superconductors by scanning SQUID microscopy (SSM)\cite{Kirtley1996} and in spin-triplet $\beta$-Bi$_2$Pd polycrystalline rings by resistance oscillations\cite{Li2019}. Indications for the presence of half-quantum vortices were also reported in micrometer size \SRO \cite{Jang2011,Cai2022}, either by cantilever magnetometry or magneto-transport, but could not be corroborated by SSM in bulk crystals\cite{Kirtley2007}.
In hole doped Ba$_{1-x}$K$_{x}$Fe$_{2}$As$_{2} (x=0.77)$, SSM detected magnetic features carrying fractional quanta of flux only at temperatures above 0.8 T/T$_{c}$\cite{Iguchi2023}. The fractional quantisation in this case is attributed to multiple bands.

The theory for domain walls between the opposite time reversed states in the B-phase of UPt$_3$ is well established \cite{Sigrist1989,Sigrist1999,Volovik2003,Ichioka2005}. It is energetically favorable to locate vortices on the domain walls compared to in the bulk, leading to the formation of vortex sheets, and for these vortices to split into pairs of fractional vortices
\cite{Sigrist1999}. The presence of domain walls has been argued to affect flux penetration and flux motion \cite{Sauls1994,Sigrist1999}. The domain walls were assumed to be immobile in this context presenting a barrier to flux motion.
While these ideas have been applied to interpret measurements of flux creep and flux noise, the direct observation of domain walls and vortex splitting are missing. In this study, we report such direct observations of half-quantum vortices and vortex sheets in the B-phase of heavy fermion UPt$_{3}$ in images taken with a scanning SQUID microscope\cite{Hykel2014, Garcia_Campos2021}. The observed features add support to the identification of TRS breaking state in the B-phase. What is less expected is that we find that the domain walls are highly mobile, moving in response to small temperature changes. 

\section*{Results}
\subsection*{Half-quantum vortices}

\begin{figure}[th]
	\centering
    \includegraphics[width=0.9\linewidth]{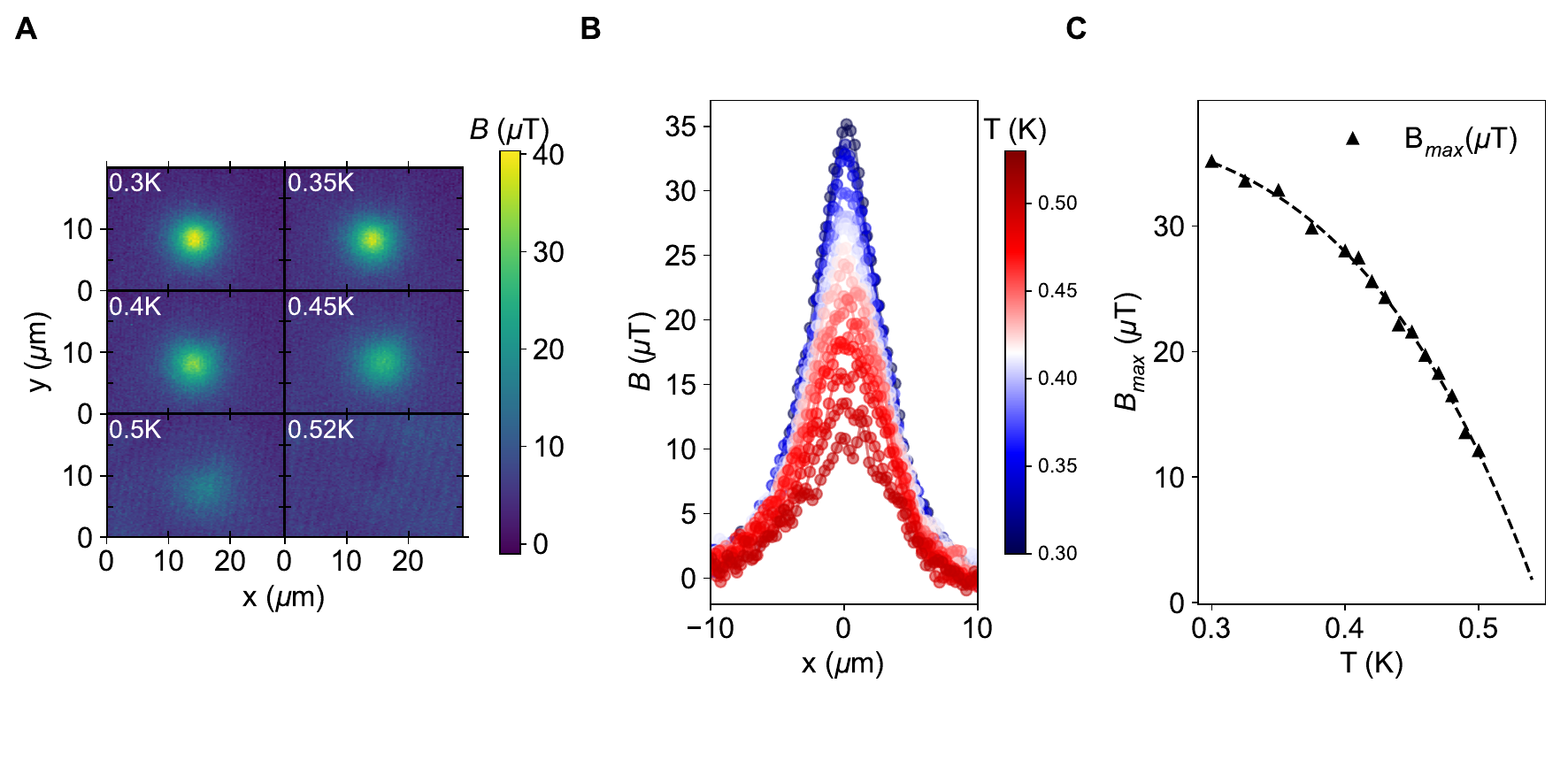}
	\caption{ \textbf{Temperature dependence of magnetic induction 
  of a single vortex in UPt$_{3}$} (A) Consecutive magnetic images of the same vortex as temperature is increased step-wise. (B) Profiles through this vortex for temperatures from 0.3 to 0.5K. (C) The temperature dependence of the maximum field at the center of the vortex. 
  }
	\label{Fig_1}
\end{figure}

Two UPt$_{3}$ single crystals from different batches (their detailed characterization is given in figure~\ref{s_Tc_ro_Dai}, \ref{s_Tc_cp_xac}, \ref{fig:s_cp_huxley}) were investigated by SSM.  A microSQUID  (figure~\ref{fig:s1}) was scanned over polished $ab$ faces of the crystals to map the emanating magnetic flux with an external magnetic field applied parallel to the crystals' $\textit{c}$-axis. After field cooling the sample in a magnetic field close to zero, flux structures were observed consistent with single vortices. A typical vortex is illustrated in Fig.~\ref{Fig_1}A, (and figure \ref{s_cut_x_cut_y}) measured for increasing temperature, starting from the low temperature B-phase up to T$_c^+$. The computation of the magnetic flux is achieved (after subtraction of a background plane) from a 2-dimensional fit to a monopole magnetic charge model  \cite{Carneiro2000}, equation \ref{eq:1}, and corresponds to a flux quantum $\phi_0$. Magnetic field profiles are displayed in panel B, from which the peak magnetic field B$_{max}$ is deduced. The profiles presented are smoothed by averaging the central point with the 8 data points around.

The temperature dependence of B$_{max}$ is shown in panel C, with B$_{max}$ = 35\,$\mu$T at 0.3\,K. 
The dashed lines represent a fit with a power law model 
B$_{max}$=B$_{0}$(1-(T/T$_{c})^{m})$, between 0.3 and 0.5 K, from which we extracted the parameters 
B$_{0}$=38 $\pm$ 1 $\mu$T, $m$=4.2 $\pm$ 0.1, and T$_{c}$=0.54 $\pm$ 0.01K 

\begin{figure}[h!]
	\centering	\includegraphics[width=0.7\linewidth]{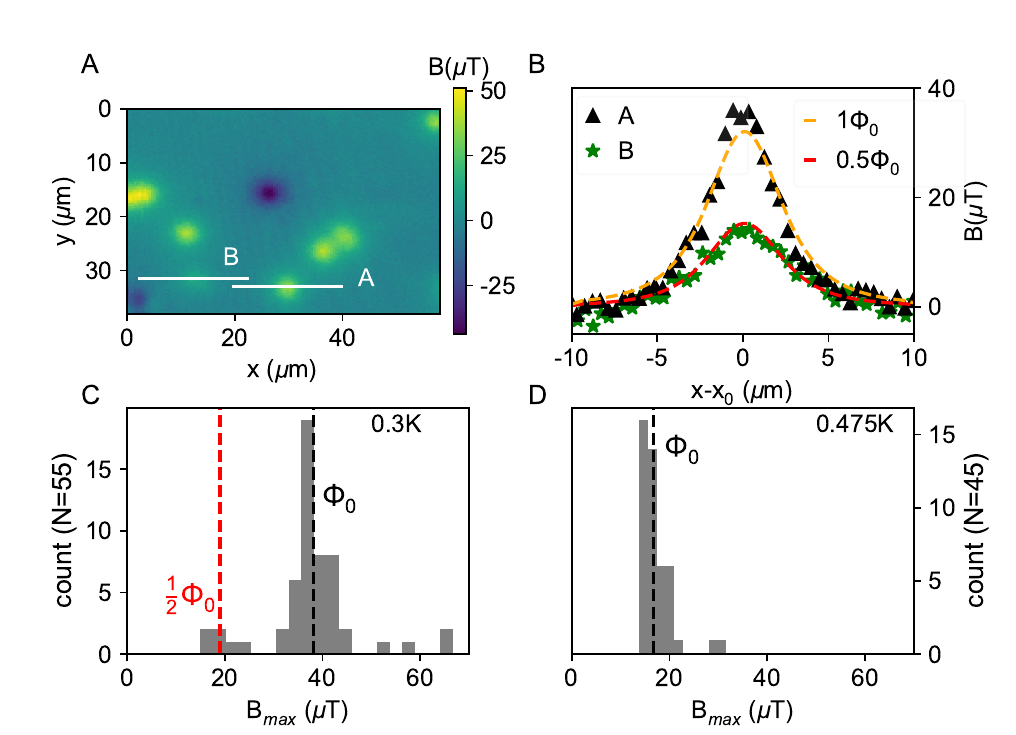}
\caption{\textbf{Vortices, antivortices and half-quantum vortices in the B-phase of \UPt.} (A) An image of the local magnetic field over a sample surface of UPt$_3$ following cooling from above $T_c^+$ to 0.3\,K in close to zero applied magnetic field.  Several vortices, two anti-vortices and a half-quantum vortex are visible. The lines labeled A and B pass through the center of a vortex and a half-quantum vortex respectively.  (B) The field profiles along the lines shown in panel (A). The solid curves are fits (see text) to these profiles with flux $\phi_0$ and $\phi_0/2$ (and the same SQUID to sample surface distance). (C) is a frequency histogram of local field maxima values from 7 such images captured in 7 consecutive cool downs from above $T_{c^+}$ to 0.3\,K in the B-phase. The vertical dashed lines show the field maximum values corresponding to 0.5 $\phi_0$ and $\phi_0$ vortices.  (D) The same procedure was followed as in (C) except that the sample was cooled instead to 0.475\,K in the A-phase. The vertical dashed line shows the $B_\text{max}$ value for full quantum $\phi_0$ vortices at this temperature; no half quantum vortices were seen.}
	\label{Fig_2}
\end{figure}

In the B-phase integer quantum vortices are not the only flux structures detected. 
Weaker signals from vortices with half of the maximum field amplitude are also visible in the magnetic image, Fig.~\ref{Fig_2}(A). The line-profiles of a 1$\Phi_{0}$ vortex (line A) and a 0.5$\Phi_{0}$ vortex (line B) are plotted in Fig.~\ref{Fig_2}(B), along with fits to the monopole model (equation \ref{eq:1}) with nearly identical parameters for penetration depth + SQUID sample distance  (with $\Phi_{0}$ as prefactor: 3.31 $\pm$ 0.02 $\mu$m and with $\Phi_{0}$/2 3.26$\pm$ 0.02 $\mu$m ).
The magnetic field profiles of both single and half quantum vortices are isotropic within the resolution of the measurements; i.e. are the same along orthogonal directions (see figure \ref{s_cut_x_cut_y}).

In order to study whether half-quantum vortices are specific to the B-phase, we acquired consecutive images after field cooling (FC), firstly in the B-phase at 0.3\,K and then after increasing the temperature to 0.475~K in the A-phase (repeated 7 times). Histograms of the frequency of the field maxima values observed in all the images at each temperature are shown in Fig.\ref{Fig_2}(C) and (D). At 0.3\,K (panel C), the strongest peak in the distribution is at 38 $\mu$T corresponding to 1$\Phi_{0}$ vortices. A second peak at half this value (19 $\mu$T) corresponds to half-quantum vortices. There are some field maxima values above 38 $\mu$T that can be assigned to overlapping vortices. At 0.475\,K (panel D), the main peak is centered at 17 $\mu$T, which correspond to 1$\Phi_{0}$ vortices at this temperature. There are no maxima with lower field values. These observations are consistent with the presence of half-quantum vortices in the B-phase, and their absence in the A-phase.

\begin{figure}[h!]
	\centering
	\includegraphics[width=0.9\linewidth]{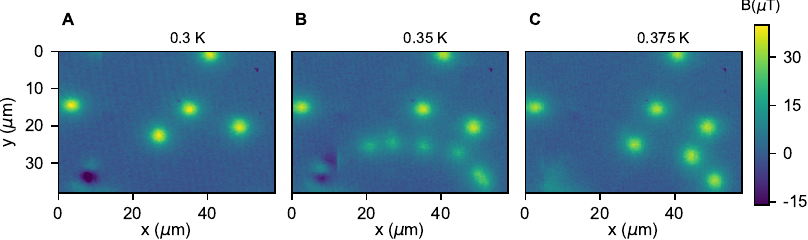}
	\caption{\textbf{Emergence and merging of half-quantum vortices in the B-phase of UPt$_3$.} (A) A magnetic image obtained after cooling from abouve $T_c^+$ in close to zero field to 0.3\,K, then (B) after subsequently increasing the temperature to 0.35\,K and (C) after further increasing the temperature to 0.375\,K. In panel (B) the single 1$\Phi_{0}$ vortex situated close to the center of the image in panel (A) has split into two half-quantum vortices and four other half-quantum vortices have entered the imaged region in the lower right quadrant forming an arc of half-quantum vortices. In panel (C) the half quantum vortices have combined to leave only full quantum vortices. Further details are discussed in the main text.}
	\label{Fig_3}
\end{figure}

To investigate the stability of the half-quantum vortices in the B-phase the sample was cooled to 0.3\,K in 
a field compensating the laboratory residual magnetic field, and images were acquired during 1 hour at various temperatures in 25\,mK steps without surpassing the set temperature by more than 1 mK. The image at 0.3\,K (Fig.\ref{Fig_3}A), contains five clear quantum vortices plus a vortex/antivortex structure in the lower left corner. We discuss this last feature separately after first discussing the main part of the image containing the 5 vortices. 
At 0.35\,K (Fig.\ref{Fig_3}B), in addition to the four vortices situated in the upper half of the image whose positions are unchanged (pinned), several half-quantum vortices appear in the lower half of the image. The fifth quantum vortex, observed at 0.3\,K close to the image center, seems to have split into two of these half-vortices, with additional half-quantum vortices entering the image from the right to form an arc of half quantum vortices.  At 0.375\,K (panel C), all the half-quantum vortices have disappeared from the imaged region and only 1$\Phi_{0}$ vortices are observed. The fifth vortex observed at 0.3\,K has reappeared in the same central location (x=25$\mu$m, y= 22 $\mu$m); it looks like the two half-quantum vortices formed from the 5th vortex have therefore recombined at 0.4~K. The other two pairs of half quantum vortices introduced at 0.35~K appear to have coalesced to leave behind two new single quantum vortices. 

The interpretation of the structure in the lower left region is less clear cut. The image at 0.3\,K may represent a very small bubble of the minority chirality containing an anti-vortex surrounded by a domain wall decorated with two half quantum vortices. The bubble is burst when threaded by a domain wall crossing the image at 0.35~K; the measurement in this region may also be affected by the field from the scanning SQUID but the image appears to comprise two pairs of half quantum vortices/anti-vortices lying on the continuation of the arc along which the other half quantum vortices are located. These combine to leave no vortices in image C. 



\subsection*{Flux domains}

To better characterize the domain structure in UPt$_3$, we have recorded images of the magnetic field distribution in different higher applied magnetic fields at different temperatures. In a conventional type-II superconductor, in the mixed state above the first critical field, a regular vortex lattice is expected and for UPt$_3$ regular vortex lattices have been observed with neutron diffraction in magnetic fields higher than 100\,mT/$\mu_{0}$\cite{Huxley2000,Avers2020}. Our measurements are at much lower applied fields of less than 1~mT.


\begin{figure}[h!]
	
	\centering
	\includegraphics[width=0.7\linewidth]{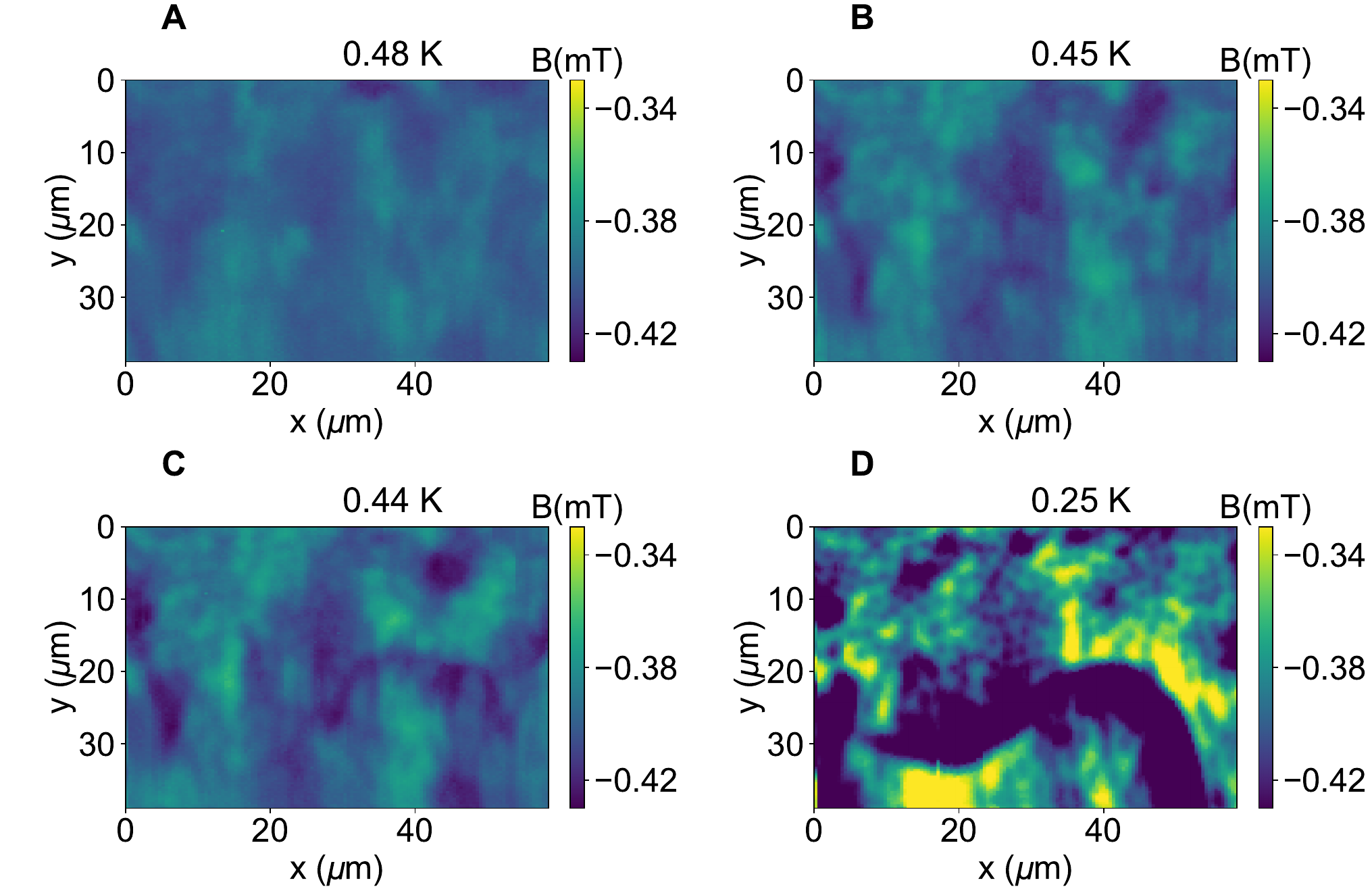}
	\caption{\textbf{Domain walls in the B-phase of \UPt}. The images were acquired when decreasing the temperature (from the A-phase to the B-phase) in a -0.4\,mT/$\mu_{0}$ applied magnetic field. In the B-phase, between 0.45\,K and 0.44\,K, an extended domain wall first starts to become discernible. The contrast across the domain wall becomes very intense at 0.25\,K and single vortices are visible on either side of the domain wall (a cut through the data in panel D is shown in Fig. S5). 
		}
	\label{Fig_4}
\end{figure}
 
Fig.~\ref{Fig_4} shows the evolution of the magnetic field during a cool down from the A- to the B-phase under an applied magnetic field of -0.4\,mT/$\mu_{0}$. 
In the A-phase, we observe an irregular arrangement of flux, probably due to the softening of inter-vortex forces and weakening of the alignment of the vortices with the applied field direction, both owing to the divergence of the penetration depth as temperature approaches the superconducting-normal transition. 
The symmetry breaking field (SBF), commonly assumed to arise from static anti-ferromagnetism, is known to have a correlation length as short as 300 \AA \cite{vanDijk1998}, which is smaller than the penetration depth. A uniform or longer range component of the SBF is required to explain the sharp thermodynamic transitions at $T_c^\pm$ in the heat capacity \cite{Graf2001}. The short range disorder of the SBF may however contribute to the inhomogeneity of the vortex distribution in the A-phase.

Further cooling into the B-phase results in continuous lines of flux contrast forming below 0.440\,K (Fig.~\ref{Fig_4}C) extending along lengths of at least 40 $\mu$m, which becomes much more intense at 0.25\,K. Above and below these lines, individual vortices are clearly visible (see also figure~\ref{s_zoom_vortex_vortex_sheet}).
The lines are present only in the B-phase and correspond well with predictions for domain walls between opposite chirality superconducting states. 
In Fig.~\ref{Fig_4}D, the average flux captured by the SQUID over the whole image is close to the applied -0.4\,mT/$\mu_{0}$ value. 

\begin{figure}[h!]
	\centering
	\includegraphics[width=0.7\linewidth]{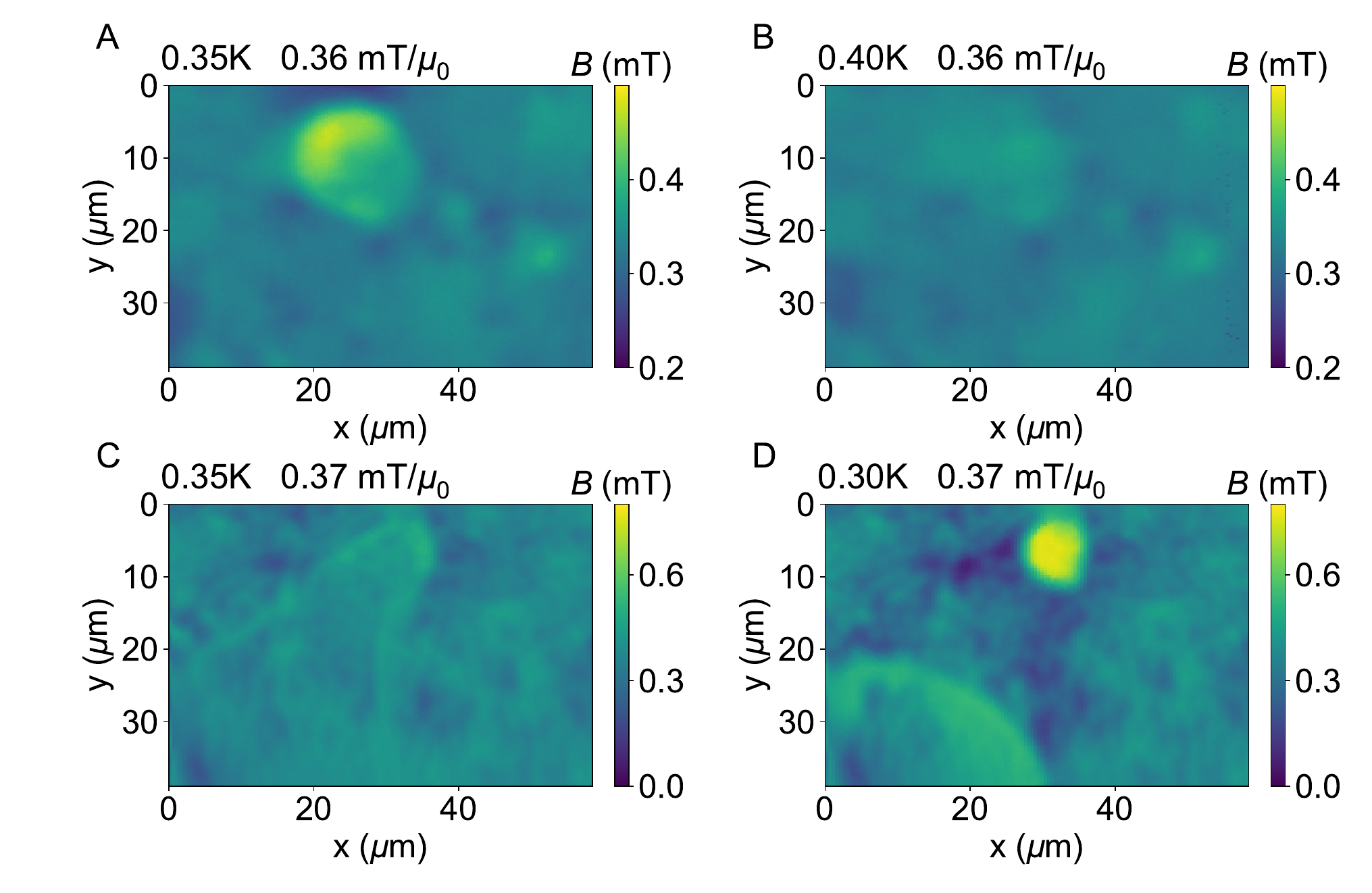}
	\caption{\textbf{Temperature evolution of domain structure}. The figure compares consecutive magnetic images for warming and cooling in a small applied field.   (A) is the magnetic image following cooling from above $T_c^+$ to 0.35\,K in an applied field 0.37\,mT/$\mu_{0}$. (B) is the image after subsequently warming the sample to 0.4\,K. (C) Is the image following cooling from above $T_c^+$ to 0.35 \,K in 0.36\,mT/$\mu_{0}$. (D) is the image following subsequently cooling the sample further to 0.30\,K. Circular domains are found in images (A) and (D). A region of reduced flux density is observed in (D) at the position occupied by the domain wall in (C). 
	}
	\label{Fig_5}
\end{figure}

The shape of the domains enclosed by the lines of contrast evolve as a function of temperature. To determine how the flux distribution changes with temperature, we first field cooled the sample in 0.36\,mT/$\mu_{0}$ to 0.35\,K and after a first image was acquired, increased the temperature to 0.4\,K and recorded a second image (warming cycle). Subsequently, the sample was heated above T$_c$ and field cooled again in 0.37\,mT/$\mu_{0}$ to 0.35\,K measured and then cooled down to 0.3\,K and measured again (cooling cycle). The results are displayed in Fig. \ref{Fig_5}. At higher temperatures, domain walls seem to be more extended (Fig. \ref{Fig_5}B-C), while at lower temperatures the walls enclose circular domains squeezing the flux density enclosed. Fig. \ref{Fig_5}D shows how a state with an extended domain structure, with an almost uniform flux distribution, evolves into a state dominated by two circular structures on cooling. It appears that the domain walls shorten with decreasing temperature. The dominant feature of the magnetic field profile crossing a domain wall is the depletion of flux on the exterior edge and an accumulation of flux on the interior edge visible in Fig. 4D.  Detailed field profiles crossing a domain wall (Fig. S5) confirm this.
This is consistent with the walls pushing and pulling on the neighboring flux lines when they move. When the sample is cooled quickly a region of reduced vortex density appears to have been left along a line vacated by a wall, seen in Fig 5D.



The same type of domain walls/vortex sheets were observed in a second sample, with similar domain shapes following field cooling (Figs.~\ref{fig:s_pos_Dolocan},~\ref{fig:s_pos_neg_Dolocan}). The contrast at the domain walls is stronger in regions of high curvature, where the wall tension might be largest opposing a stronger field gradient across the wall (Fig.~\ref{fig:s_pos_neg_Dolocan}D). These measurements also showed that the circular form of the domain walls does not depend on the sense of the field (positive or negative).  After each cool down the domain walls nucleate at different locations.

\section*{Discussion}


Our results show clear evidence of half-quantum vortices 
and vortex sheet formation in single crystals of the heavy fermion compound UPt$_3$. 
These features are present once the sample is cooled in a magnetic field between zero and $\sim$1 mT/$\mu_{0}$ parallel to the c-axis in the B phase, but not in the A-phase. In this context it has already been established that the Meissner effect is very weak (at best 2-3\% of diamagnetic shielding at 0.5\,mT/$\mu_{0}$ \cite{wuchner_magnetic_1993,Vincent1991}). 
The sheets are continuous, and define different superconducting domains. The presence of domains with opposite chirality offer a compelling explanation of these observations. 
It is however perhaps surprising that domains are present at all since domain walls cost energy. For a TRS broken state, a planar sample (as studied) would naturally contain domains owing to a demagnetizing field. Domains in the B-phase may also result from orientation domains in the A-phase from which the B-phase forms. 

The contrast at the walls is strongly correlated with the wall curvature. The magnitude of the field is consistently higher on the interior side of the wall. There is no clear evidence for additional contributions arising from chiral currents that would also be present across straight sections of wall \cite{Sato2017};
our measurement resolution in small applied fields is insufficient to detect the presence of chiral currents of even the largest magnitude predicted by theory, relative to the field produced from nearby vortices (see SM). The observed magnetic field profiles when approaching domain walls are dominated by flux pinning, resulting in field gradients extending over more macroscopic lengths of $\sim 10~\mu \text{m}$ compared to the penetration depth (the length scale over which chiral currents are screened). 

The domain wall surface energy $\gamma$ can be estimated from the field profile across the wall and radius of curvature $R$ of the domain wall alone. It is $\gamma= R~ \Delta P$ with $\Delta P$ the pressure across the wall due to the Lorentz force acting between the current density (determined from the field gradient across the wall) and the flux in the wall. The current density is calculated assuming the measured field change across the wall occurs over a penetration depth (see SM). The data in Fig 4 (cuts through the data are shown in Fig. S5) gives $\gamma_\text{expt.} = 5\times 10^{-6}~\text{Jm}^{-2}$ at 0.25\,K and 0.4 mT. This value can be compared to a theoretical estimate based on the Ginzburg-Landau and microscopic theory. The theoretical estimate for $\gamma$ is determined from the loss of condensation energy density due to the suppression of the order parameter at the wall times the distance perpendicular to the wall over which the suppression occurs. For a BCS superconductor the condensation energy density is $\frac{1}{2} g_\uparrow(0) |\Delta|^2$ with $g_\uparrow(0)$ the density of states per volume per spin at the Fermi level and $|\Delta|$ the magnitude of the gap with $|\Delta|\sim 1.76 k_BT_c$ for $T/T_c\lesssim 0.5$ (the BCS constant 1.76 is modified only slightly for a non-conventional superconductor with $|\Delta|$ replaced by the gap averaged over the Fermi surface \cite{sigrist_introduction_2005}). At the domain wall in \UPt only one component of the two component order parameter is suppressed; the energy cost is therefore reduced by around 65\% (see SM) compared to fully suppressing the superconductivity. The gradient energy and loss of condensation energy give equal contributions to the wall energy in the absence of gradient coupling terms. The density of states is obtained directly from the normal state specific heat coefficient. Inserting these values and setting the distance over which the condensation energy is suppressed equal to the coherence length $\xi =110 \text{\AA}$ gives a theoretical wall energy $\gamma_\text{theory} \sim 5\times 10^{-6} ~\text{Jm}^{-2}$. The good agreement between the experimental and the theoretical estimates of the wall energy support the identification that the walls are boundaries  between anti-phase superconducting domains.


What is clear from our measurements is that the domain wall energy and surface tension increase strongly as the temperature is reduced below $T_c^-$  causing the domain walls to shrink, compressing the flux on their interior side and leaving a lower flux density on their exterior side, over a typical length scale of 10 $\mu$m. This dominates any field difference across the walls that might be due to edge currents.
The large displacements of the domain walls in response to changes of temperature contrast with fixed positions of vortices in the bulk. At low temperature we observed that the minority domains contract into almost circular bubbles in agreement with predictions from Landau-Ginzburg theory \cite{Matsunaga2004}.
These bubbles survive down to the lowest temperature accessible in our experiments of 0.25~K.  It would be interesting to determine whether domains and bubbles are also present at higher fields than those we examined.

In very low field and at 350 mK we were able to resolve individual half quantum vortices along an arc 
as predicted to occur along chiral domain walls \cite{Sigrist1989,Volovik2003, Matsunaga2004,Ichioka2005,Etter2020}. This provides clear additional evidence that the B-phase breaks TRS.  The background field in this case is small with vortices / half quantum vortices widely spaced. This provides a much smoother background to look for the presence of chiral currents.  
The maximum theoretical magnitude of the chiral current for the most favorable choice of order parameter would give a step of $8 \mu\text{T}$ across the wall for a surface-SQUID distance of 1.8$\mu$m (the calculation and a simulated profile is given in the SM and Fig S10B).  Clearly no such step is seen in the measurements with an experimental resolution of $\pm 1.5 \mu \text{T}$ (the field profile measured along different lines crossing the domain wall in Fig 3B are shown in Fig. S9). Bulk chiral currents much smaller than the maximum value estimate are expected for Saul's $E_{2u}$ state. The detected field step would also be reduced for all order parameters if the chiral currents did not extend to the sample surface. Thus, while our measurements place a constraint on the magnitude of currents extending to the surface, they do not constrain how large any chiral currents may be in the bulk and therefore do not rule out different order parameter choices that are otherwise compatible with chiral domains.

In future studies it would be interesting to pattern the crystals with a focused ion beam \cite{Bachmann2019} to create preferential pinning sites for both the walls and vortices, as well as to examine how the vortices and walls respond to electrical and/or heat currents. The presence of half-quantum vortices on the domain walls, makes UPt$_{3}$ a strong candidate for hosting Majorana zero modes \cite{Kallin2016}.  We hope our measurements will motivate scanning tunneling microscopy and other measurements that can directly probe the excitation spectra in the half quantum vortices to look more directly for zero bias resonances from such modes, as well as further theoretical work.

\pagebreak
\newpage
\clearpage
\bibliography{UPt3_Bib_nett}

@article{Volovik1976,
    title = {Line and point singularities in superfluid $^3He$},
    author = {Volovik, G. E. and Mineev, V. P.},
    journal = {Jetp Letters},
    volume = {24},
    issue = {11},
    pages = {605},
    year = {1976},

}

@inproceedings{sigrist_introduction_2005,
	address_shit = {Salerno (Italy)},
	title = {Introduction to {Unconventional} {Superconductivity}},
	volume = {789},
	url_shit = {https://pubs.aip.org/aip/acp/article/789/1/165-243/692377},
	doi_shit = {10.1063/1.2080350},
	language = {en},
	urldate_shit = {2025-09-04},
	booktitle = {{AIP} {Conference} {Proceedings}},
	publisher = {AIP},
	author = {Sigrist, Manfred},
	year = {2005},
	note_shit = {ISSN: 0094243X},
	pages = {165--243},
	file_shit = {(Sigrist) Introduction to unconventional superconductivity:/home/hasselbach-ubuntu/snap/zotero-snap/common/Zotero/storage/44J4N529/(Sigrist) Introduction to unconventional superconductivity.pdf:application/pdf},
}

@Article{Benfenati2020,
  author   = {Benfenati, A. and Barkman, M. and Winyard, T. and Wormald, A. Speight, M. and Babaev, E.},
  journal  = {Phys. Rev. B},
  title    = {Magnetic signatures of domain walls in s+is and s+id superconductors: Observability and what that can tell us about the superconducting order paramter},
  year     = {2020},
  month    = {},
  note_shit     = {Publisher: APS},
  pages    = {054507},
  volume   = {101},
  doi_shit        = {},
  shit_shit_url      = {},
}

@Article{Huxley2000,
  author   = {Huxley, A. and Rodi\'ere, P. and Mc Paul, D. and van Dijk, N. and Cubitt, R. and Flouquet, J.},
  journal  = {Nature},
  title    = {Realignment of the flux-line lattice by a change in the symmetry of the superconductivity in UPt$_3$},
  year     = {2000},
  month    = {},
  note_shit     = {Publisher: McMillan},
  pages    = {160},
  volume   = {406},
  doi_shit        = {},
  shit_shit_url      = {},
}

@Article{Yaouanc1998,
  author   = {Yaouanc, A. and Dalmas de R\'eotier, P. and Huxley, A. and Flouquet, J. and Bonville, P. and Gubbens, P. C. M. and Mulders, A. M.},
  journal  = {J. Phys. Condens. Matter},
  title    = {Strong axial anisotropy of the magnetic penetration length superconducting UPt$_3$},
  year     = {1998},
  month    = {},
  note_shit     = {Publisher: IOP},
  pages    = {9791},
  volume   = {10},
  doi_shit        = {},
  shit_shit_url      = {},
}

@Article{vanDijk1998,
  author   = {van Dijk, N. H. and F\r{a}k, B. and Regnault, L. P. and Huxley, A. and Fern\'andez-D\'iaz, M-T.},
  journal  = {Phys. Rev. B},
  title    = {Magnetic order of UPt$_3$ in high Magnetic fields},
  year     = {1998},
  month    = {},
  note_shit     = {Publisher: APS},
  pages    = {3186},
  volume   = {58},
  doi_shit        = {},
  shit_shit_url      = {},
}

@Article{Graf2001,
  author   = {Graf, M. and Hess, D. W.},
  journal  = {Phys. Rev. B},
  title    = {Antiferromagnetic domains and superconductivity in UPt$_3$},
  year     = {2001},
  month_shit    = {},
  note_shit     = {Publisher: APS},
  pages    = {134502},
  volume   = {63},
  doi_shit        = {},
  shit_shit_url      = {},
}

@Article{Sato2017,
  author   = {Sato, M. and Ando, Y.},
  journal  = {Reports on Progress in Physics},
  title    = {Topological superconductors: a review},
  year     = {2017},
  month    = May,
  note_shit     = {Publisher: IOP Science},
  pages    = {076501},
  volume   = {80},
  doi_shit        = { 10.1088/1361-6633/aa6ac7 SMASH},
  shit_shit_url      = {https://iopscience.iop.org/article/10.1088/1361-6633/aa6ac7},
}

@Article{Gannon2015,
  author   = {Gannon, W. J. and Halperin, W. P. and Rastovski, C. and Schlesinger, K. J. and Hlevyack, J and Eskildsen,M. R. and Vorontsov, A. B. and Gavilano, J. and Gasser, U. and Nagy, G.},
  journal  = {New Journal of Physics},
  title    = {Nodal gap structure and order parameter symmetry of the unconventional superconductor {UPt$_{3}$} },
  year     = {2015},
  month    = feb,
  note_shit     = {Publisher: IOP Science},
  pages    = {023041},
  volume   = {17},
  doi_shit        = {10.1088/1367-2630/17/2/023041},
  shit_shit_url      = {https://iopscience.iop.org/article/10.1088/1367-2630/17/2/023041},
}

@Article{Kirtley2007,
  author   = {Kirtley, J. R. and Kallin, C. and Hicks, C. W. and Kim, E.-A. and Liu, Y. and Moler, K. A. and Maeno, Y. and Nelson, K. D.},
  journal  = {Physical Review B},
  title    = {Upper limit on spontaneous supercurrents in {Sr$_{2}$RuO$_{4}$} },
  year     = {2007},
  month    = jul,
  note_shit     = {Publisher: American Physical Society},
 number_shit  = {1},
  pages    = {014526},
  volume   = {76},
  doi_shit        = {10.1103/PhysRevB.76.014526},
  shit_shit_url      = {https://link.aps.org/doi/10.1103/PhysRevB.76.014526},
  shit_shit_urldate  = {2021-07-21},
}

@Article{Adenwalla1990,
  author   = {Adenwalla, S. and Lin, S. W. and Ran, Q. Z. and Zhao, Z. and Ketterson, J. B. and Sauls, J. A. and Taillefer, L. and Hinks, D. G. and Levy, M. and Sarma, Bimal K.},
  journal  = {Physical Review Letters},
  title    = {Phase diagram of {UPt$_{3}$} from ultrasonic velocity measurements},
  year     = {1990},
  month    = oct,
  note_shit     = {Number: 18},
 number_shit  = {18},
  pages    = {2298--2301},
  volume   = {65},
  doi_shit        = {10.1103/PhysRevLett.65.2298},
  shit_shit_url      = {https://link.aps.org/doi/10.1103/PhysRevLett.65.2298},
  shit_shit_urldate  = {2020-01-07},
}

@Article{Bruls1990,
  author   = {Bruls, G. and Weber, D. and Wolf, B. and Thalmeier, P. and Lüthi, B. and Visser, A. de and Menovsky, A.},
  journal  = {Physical Review Letters},
  title    = {Strain--order-parameter coupling and phase diagrams in superconducting {UPt$_{3}$}	},
  year     = {1990},
  month    = oct,
  note_shit     = {Number: 18},
 number_shit  = {18},
  pages    = {2294--2297},
  volume   = {65},
  doi_shit        = {10.1103/PhysRevLett.65.2294},
  shit_shit_url      = {https://link.aps.org/doi/10.1103/PhysRevLett.65.2294},
  shit_shit_urldate  = {2020-01-07},
}

@Article{Schemm2014,
  author    = {Schemm, E. R. and Gannon, W. J. and Wishne, C. M. and Halperin, W. P. and Kapitulnik, A.},
  journal   = {Science},
  title     = {Observation of broken time-reversal symmetry in the heavy-fermion superconductor {UPt$_{3}$}	},
  year      = {2014},
  issn      = {0036-8075, 1095-9203},
  month     = jul,
  note_shit      = {Number: 6193},
 number_shit   = {6193},
  pages     = {190--193},
  volume    = {345},
  copyright = {Copyright © 2014, American Association for the Advancement of Science},
  doi_shit         = {10.1126/science.1248552},
  language  = {en},
  pmid      = {25013069},
  shit_shit_url       = {http://science.sciencemag.org/content/345/6193/190},
  shit_shit_urldate   = {2019-04-02},
}

@Article{Sauls1994,
  author  = {Sauls, J.A.},
  journal = {Advances in Physics},
  title   = {The order parameter for the superconducting phases of {UPt$_3$}},
  year    = {1994},
  note_shit    = {Number: 1},
 number_shit = {1},
  pages   = {113--141},
  volume  = {43},
 	doi_shit    = {10.1080/00018739400101475},
  shit_shit_url     = {http://dx.doi.org/10.1080/00018739400101475},
}

@Article{Joynt2002,
  author   = {Joynt, Robert and Taillefer, Louis},
  journal  = {Reviews of Modern Physics},
  title    = {The superconducting phases of UPt$_{3}$},
  year     = {2002},
  month    = mar,
  note_shit     = {Number: 1},
 number_shit  = {1},
  pages    = {235--294},
  volume   = {74},
  doi_shit        = {10.1103/RevModPhys.74.235},
  shit_shit_url      = {https://link.aps.org/doi/10.1103/RevModPhys.74.235},
  shit_shit_urldate  = {2019-10-02},
}

@Article{Luke1993,
  author   = {Luke, G. M. and Keren, A. and Le, L. P. and Wu, W. D. and Uemura, Y. J. and Bonn, D. A. and Taillefer, L. and Garrett, J. D.},
  journal  = {Physical Review Letters},
  title    = {Muon spin relaxation in {UPt$_{3}$} },
  year     = {1993},
  month    = aug,
  note_shit     = {Publisher: American Physical Society},
 number_shit  = {9},
  pages    = {1466--1469},
  volume   = {71},
  doi_shit        = {10.1103/PhysRevLett.71.1466},
  shit_shit_url      = {https://link.aps.org/doi/10.1103/PhysRevLett.71.1466},
  shit_shit_urldate  = {2021-03-12},
}

@Article{Strand2009,
  author   = {Strand, J. D. and Van Harlingen, D. J. and Kycia, J. B. and Halperin, W. P.},
  journal  = {Physical Review Letters},
  title    = {Evidence for {Complex} {Superconducting} {Order} {Parameter} {Symmetry} in the {Low}-{Temperature} {Phase} of {UPt$_3$} from {Josephson} {Interferometry}},
  year     = {2009},
  month    = nov,
  note_shit     = {Number: 19},
 number_shit  = {19},
  pages    = {197002},
  volume   = {103},
  doi_shit        = {10.1103/PhysRevLett.103.197002},
  shit_shit_url      = {https://link.aps.org/doi/10.1103/PhysRevLett.103.197002},
  shit_shit_urldate  = {2020-01-29},
}

@Article{Avers2020,
  author   = {Avers, K. E. and Gannon, W. J. and Kuhn, S. J. and Halperin, W. P. and Sauls, J. A. and DeBeer-Schmitt, L. and Dewhurst, C. D. and Gavilano, J. and Nagy, G. and Gasser, U. and Eskildsen, M. R.},
  journal  = {Nature Physics},
  title    = {Broken time-reversal symmetry in the topological superconductor {UPt$_3$}},
  year     = {2020},
  issn     = {1745-2473, 1745-2481},
  month    = may,
 number_shit  = {5},
  pages    = {531--535},
  volume   = {16},
  doi_shit        = {10.1038/s41567-020-0822-z},
  language = {en},
  shit_shit_url      = {http://www.nature.com/articles/s41567-020-0822-z},
  shit_shit_urldate  = {2021-03-08},
}

@Article{Reotier1995,
  author   = {de Réotier, P. Dalmas and Huxley, A. and Yaouanc, A. and Flouquet, J. and Bonville, P. and Imbert, P. and Pari, P. and Gubbens, P. C. M. and Mulders, A. M.},
  journal  = {Physics Letters A},
  title    = {Absence of zero field muon spin relaxation induced by superconductivity in the {B} phase of {UPt$_{3}$}	},
  year     = {1995},
  issn     = {0375-9601},
  month    = sep,
 number_shit  = {2},
  pages    = {239--243},
  volume   = {205},
  doi_shit        = {10.1016/0375-9601(95)00548-H},
  language = {en},
  shit_shit_url      = {https://www.sciencedirect.com/science/article/pii/037596019500548H},
  shit_shit_urldate  = {2022-01-27},
}

@Article{Nomoto2016,
  author   = {Nomoto, Takuya and Ikeda, Hiroaki},
  journal  = {Physical Review Letters},
  title    = {Exotic {Multigap} {Structure} in {UPt$_3$} {Unveiled} by a {First}-{Principles} {Analysis}},
  year     = {2016},
  month    = nov,
  note_shit     = {Publisher: American Physical Society},
 number_shit  = {21},
  pages    = {217002},
  volume   = {117},
  doi_shit        = {10.1103/PhysRevLett.117.217002},
  shit_shit_url      = {https://link.aps.org/doi/10.1103/PhysRevLett.117.217002},
  shit_shit_urldate  = {2022-01-31},
}

@article{Yanase_mobius_2017,
	title = {M{\"o}bius topological superconductivity in  {UPt$_{3}$}},
	volume = {95},
	url-shit = {https://link.aps.org/doi/10.1103/PhysRevB.95.224514},
	doi-shit = {10.1103/PhysRevB.95.224514},
	number = {22},
	urldate_shit = {2019-04-07},
	journal = {Physical Review B},
	author = {Yanase, Youichi and Shiozaki, Ken},
	month = jun,
	year = {2017},
	note-shit = {Number: 22},
	pages = {224514},
	file-shit = {APS Snapshot:/home/hasselbach-ubuntu/snap/zotero-snap/common/Zotero/storage/EEZX846Z/PhysRevB.95.html:text/html;Full Text PDF:/home/hasselbach-ubuntu/snap/zotero-snap/common/Zotero/storage/SZ5A8L2G/Yanase et Shiozaki - 2017 - Mobius topological superconductivity in \$ mathr.pdf:application/pdf},
}

@Article{Yanase2016,
  author   = {Yanase, Youichi},
  journal  = {Physical Review B},
  title    = {Nonsymmorphic {Weyl} superconductivity in {UPt$_3$} based on {E$_{2u}$} representation},
  year     = {2016},
  month    = nov,
  note_shit     = {Publisher: American Physical Society},
 number_shit  = {17},
  pages    = {174502},
  volume   = {94},
  doi_shit        = {10.1103/PhysRevB.94.174502},
  shit_shit_url      = {https://link.aps.org/doi/10.1103/PhysRevB.94.174502},
  shit_shit_urldate  = {2021-10-19},
}

@Article{Sigrist1989,
  author   = {Sigrist, M. and Rice, T. M. and Ueda, K.},
  journal  = {Physical Review Letters},
  title    = {Low-field magnetic response of complex superconductors},
  year     = {1989},
  month    = oct,
  note_shit     = {Publisher: American Physical Society},
 number_shit  = {16},
  pages    = {1727--1730},
  volume   = {63},
  doi_shit        = {10.1103/PhysRevLett.63.1727},
  shit_shit_url      = {https://link.aps.org/doi/10.1103/PhysRevLett.63.1727},
  shit_shit_urldate  = {2021-03-08},
}

@Article{Sigrist1999,
  author  = {Sigrist, M. and Agterberg, D. F.},
  journal = {Prog. Theor. Phys},
  title   = {The {Role} of domain {Walls} on the {Vortex} {Creep} {Dynamics} in {Unconventional} {Superconductors}},
  year    = {1999},
  pages   = {965},
  volume  = {102},
}

@Article{Ichioka2005,
  author   = {Ichioka, Masanori and Matsunaga, Yasushi and Machida, Kazushige},
  journal  = {Physical Review B},
  title    = {Magnetization process in a chiral p-wave superconductor with multidomains},
  year     = {2005},
  month    = may,
  note_shit     = {Publisher: American Physical Society},
 number_shit  = {17},
  pages    = {172510},
  volume   = {71},
  doi_shit        = {10.1103/PhysRevB.71.172510},
  shit_shit_url      = {https://link.aps.org/doi/10.1103/PhysRevB.71.172510},
  shit_shit_urldate  = {2021-03-08},
}

@Book{Volovik2003,
  author       = {{Grigory} {E} Volovik},
  publisher    = {Clarendon Press Oxford},
  title        = {The Universe in a Helium Droplet},
  year         = {2003},
  howpublished = {Clarendon Press},
  shit_shit_url          = {https://b-ok.cc/book/454292/879444},
  shit_shit_urldate      = {2019-12-05},
}

@Article{Hykel2014,
  author  = {Hykel, D. J. and Wang, Z. S. and Castellazzi, P. and Crozes, T. and Shaw, G. and Schuster, K. and Hasselbach, K.},
  journal = {Journal of Low Temperature Physics},
  title   = {{MicroSQUID} {Force} {Microscopy} in a {Dilution} {Refrigerator}},
  year    = {2014},
  pages   = {861--867},
  volume  = {175},
  annote_shit  = {Journal of Low Temperature Physics {\textbackslash}textbackslashcopyright Springer Science+Business Media New York 2014 10.1007/s10909-014-1174-9 MicroSQUID Force Microscopy in a Dilution Refrigerator D. J. Hykel1, Z. S. Wang1, 2, P. Castellazzi1, T. Crozes1, G. Shaw1, K. Schuster3 and K. Hasselbach1},
}

@Article{Etter2020,
  author  = {Etter, Sarah B and Huang, Wen and Sigrist, Manfred},
  journal = {New Journal of Physics},
  title   = {Half-quantum vortices on \textit{c} -axis domain walls in chiral p-wave superconductors},
  year    = {2020},
  issn    = {1367-2630},
  month   = sep,
 number_shit = {9},
  pages   = {093038},
  volume  = {22},
 	doi_shit    = {10.1088/1367-2630/abafe8},
  shit_shit_url     = {https://iopscience.iop.org/article/10.1088/1367-2630/abafe8},
  shit_shit_urldate = {2021-10-19},
}

@PhdThesis{Dolocan2005,
  author  = {Dolocan, V. O.},
  school  = {Université Joseph Fourier Grenoble 1},
  title   = {Imagerie magnétique des supraconducteurs non conventionnels},
  year    = {2005},
  address = {CNRS CRTBT Grenoble},
  month   = dec,
}

@PhdThesis{Garcia_Campos2021,
  author  = {Garc\'ia-Campos, P.},
  school  = {Université Grenoble Alpes},
  title   = {Visualization of Chiral Superconductivity in UPt$_{3}$},
  year    = {2021},
  address = {Institut Néel, Grenoble},
  month   = dec,
}

@Article{Matsunaga2004,
  author   = {Matsunaga, Yasushi and Ichioka, Masanori and Machida, Kazushige},
  journal  = {Physical Review B},
  title    = {Flux flow and pinning of the vortex sheet structure in a two-component superconductor},
  year     = {2004},
  month    = sep,
  note_shit     = {Publisher: American Physical Society},
 number_shit  = {10},
  pages    = {100502},
  volume   = {70},
  doi_shit        = {10.1103/PhysRevB.70.100502},
  shit_shit_url      = {https://link.aps.org/doi/10.1103/PhysRevB.70.100502},
  shit_shit_urldate  = {2022-02-23},
}

@Article{Vincent1991,
  author   = {Vincent, E. and Hammann, J. and Taillefer, L. and Behnia, K. and Keller, N. and Flouquet, J.},
  journal  = {Journal of Physics: Condensed Matter},
  title    = {Low-field diamagnetic response of the superconducting phases in {UPt$_{3}$}	},
  year     = {1991},
  issn     = {0953-8984},
  month    = may,
  note_shit     = {Publisher: IOP Publishing},
 number_shit  = {20},
  pages    = {3517--3525},
  volume   = {3},
  doi_shit        = {10.1088/0953-8984/3/20/013},
  language = {en},
  shit_shit_url      = {https://doi.org/10.1088/0953-8984/3/20/013},
  shit_shit_urldate  = {2021-07-27},
}

@Article{Li2019,
  author   = {Li, Yufan and Xu, Xiaoying and Lee, M.-H. and Chu, M.-W. and Chien, C. L.},
  title    = {Observation of half-quantum flux in the unconventional superconductor {$\beta$}-{Bi$_2$Pd}},
  journal  = {Science},
  year     = {2019},
  volume   = {366},
 number_shit  = {6462},
  pages    = {238--241},
  month    = oct,
  issn     = {0036-8075, 1095-9203},
  doi_shit        = {10.1126/science.aau6539},
  language = {en},
  shit_shit_url      = {https://www.science.org/doi/10.1126/science.aau6539},
  shit_urldate  = {2024-07-29},
}

@Article{Autti2016,
  author   = {Autti, S. and Dmitriev, V. V. and Makinen, J. T. and Soldatov, A. A. and Volovik, G. E. and Yudin, A. N. and Zavjalov, V. V. and Eltsov, V. B.},
  title    = {Observation of {Half}-{Quantum} {Vortices} in {Topological} {Superfluid} {$^{3}$}{He}  },
  journal  = {Phys. Rev. Lett.},
  year     = {2016},
  volume   = {117},
 number_shit  = {25},
  pages    = {255301},
  month    = dec,
  issn     = {0031-9007},
  note_shit     = {Place: College Pk Publisher: Amer Physical Soc WOS:000390226400002},
  doi_shit        = {10.1103/PhysRevLett.117.255301},
  language = {English},
  shit_url      = {http://www.webofscience.com/wos/woscc/full-record/WOS:000390226400002},
  shit_urldate  = {2022-10-09},
}

@Article{Kallin2016,
  author   = {Kallin, Catherine and Berlinsky, John},
  title    = {Chiral superconductors},
  journal  = {Rep. Prog. Phys.},
  year     = {2016},
  volume   = {79},
 number_shit  = {5},
  pages    = {054502},
  month    = apr,
  issn     = {0034-4885},
  note_shit     = {Publisher: IOP Publishing},
  doi_shit        = {10.1088/0034-4885/79/5/054502},
  language = {en},
  shit_url      = {https://doi.org/10.1088%2F0034-4885%2F79%2F5%2F054502},
  shit_urldate  = {2020-10-02},
}

@Article{Parts1994,
  author   = {Parts, U. and Thuneberg, E. V. and Volovik, G. E. and Koivuniemi, J. H. and Ruutu, V. M. H. and Heinilä, M. and Karimäki, J. M. and Krusius, M.},
  title    = {Vortex sheet in rotating superfluid {$^{3}$}{He}-A	},
  journal  = {Phys. Rev. Lett.},
  year     = {1994},
  volume   = {72},
 number_shit  = {24},
  pages    = {3839--3842},
  month    = jun,
  note_shit     = {Publisher: American Physical Society},
  doi_shit        = {10.1103/PhysRevLett.72.3839},
  shit_url      = {https://link.aps.org/doi/10.1103/PhysRevLett.72.3839},
  shit_urldate  = {2022-02-23},
}

@Article{Kirtley1996,
  author   = {Kirtley, J. R. and Tsuei, C. C. and Rupp, Martin and Sun, J. Z. and Yu-Jahnes, Lock See and Gupta, A. and Ketchen, M. B. and Moler, K. A. and Bhushan, M.},
  title    = {Direct {Imaging} of {Integer} and {Half}-{Integer} {Josephson} {Vortices} in {High}- ${{T}_{c}}$ {Grain} {Boundaries}},
  journal  = {Phys. Rev. Lett.},
  year     = {1996},
  volume   = {76},
 number_shit  = {8},
  pages    = {1336--1339},
  month    = feb,
  note_shit     = {Publisher: American Physical Society},
  doi_shit        = {10.1103/PhysRevLett.76.1336},
  shit_url      = {https://link.aps.org/doi/10.1103/PhysRevLett.76.1336},
  shit_urldate  = {2024-09-17},
}

@Article{Jang2011,
  author  = {Jang, J. and Ferguson, D. G. and Vakaryuk, V. and Budakian, R. and Chung, S. B. and Goldbart, P. M. and Maeno, Y.},
  title   = {Observation of {Half}-{Height} {Magnetization} {Steps} in {Sr$_{2}$RuO$_{4}$}	},
  journal = {Science},
  year    = {2011},
  volume  = {331},
 number_shit = {6014},
  pages   = {186--188},
  month   = jan,
  note_shit    = {Publisher: American Association for the Advancement of Science},
 	doi_shit    = {10.1126/science.1193839},
  shit_url     = {http://www.science.org/doi/10.1126/science.1193839},
  shit_urldate = {2022-03-02},
}

@Article{Cai2022,
  author   = {Cai, Xinxin and Zakrzewski, Brian M. and Ying, Yiqun A. and Kee, Hae-Young and Sigrist, Manfred and Ortmann, J. Elliott and Sun, Weifeng and Mao, Zhiqiang and Liu, Ying},
  title    = {Magnetoresistance oscillation study of the spin counterflow half-quantum vortex in doubly connected mesoscopic superconducting cylinders of {Sr$_{2}$RuO$_{4}$}},
  journal  = {Phys. Rev. B},
  year     = {2022},
  volume   = {105},
 number_shit  = {22},
  pages    = {224510},
  month    = jun,
  note_shit     = {Publisher: American Physical Society},
  doi_shit        = {10.1103/PhysRevB.105.224510},
  shit_url      = {https://link.aps.org/doi/10.1103/PhysRevB.105.224510},
  shit_urldate  = {2024-09-13},
}

@Article{Iguchi2023,
  author   = {Iguchi, Yusuke and Shi, Ruby and Kihou, Kunihiro and Lee, Chul-Ho and Grinenko, Vadim and Babaev, Egor and Moler, Kathryn A.},
  title    = {Observation of superconducting vortices carrying a temperature-dependent fraction of the flux quantum},
  journal  = {Science},
  year     = {2023},
  volume   = {380},
 number_shit  = {6651},
  pages    = {1244--1247},
  month    = jun,
  issn     = {0036-8075, 1095-9203},
  note_shit     = {arXiv:2301.12368 [cond-mat]},
  doi_shit        = {10.1126/science.abp9979},
  language = {en},
  shit_url      = {http://arxiv.org/abs/2301.12368},
  shit_urldate  = {2024-02-07},
}

@Article{Matsunaga2004a,
  author   = {Matsunaga, Yasushi and Ichioka, Masanori and Machida, Kazushige},
  title    = {Vortex {State} in {Double} {Transition} {Superconductors}},
  journal  = {Phys. Rev. Lett.},
  year     = {2004},
  volume   = {92},
 number_shit  = {15},
  pages    = {157001},
  month    = apr,
  note_shit     = {Publisher: American Physical Society},
  doi_shit        = {10.1103/PhysRevLett.92.157001},
  shit_url      = {https://link.aps.org/doi/10.1103/PhysRevLett.92.157001},
  shit_urldate  = {2021-03-08},
}

@Article{Dijk1993,
  author   = {van Dijk, N. H. and de Visser, A. and Franse, J. J. M. and Holtmeier, S. and Taillefer, L. and Flouquet, J.},
  title    = {Expansivity of the superconducting phases of {UPt$_{3}$} },
  journal  = {Phys. Rev. B},
  year     = {1993},
  volume   = {48},
 number_shit  = {2},
  pages    = {1299--1302},
  month    = jul,
  note_shit     = {Publisher: American Physical Society},
  doi_shit        = {10.1103/PhysRevB.48.1299},
  shit_url      = {https://link.aps.org/doi/10.1103/PhysRevB.48.1299},
  shit_urldate  = {2024-10-16},
}

@Article{Volovik1985,
  author  = {Volovik, G. E. and Gorkov, L. P.},
  title   = {Superconducting classes in heavy fermion systems},
  journal = {Sov. Phys. JETP},
  year    = {1985},
  volume  = {61},
  pages   = {843},
}

@Article{Bachmann2019,
  author   = {Bachmann, Maja D. and Ferguson, G. M. and Theuss, Florian and Meng, Tobias and Putzke, Carsten and Helm, Toni and Shirer, K. R. and Li, You-Sheng and Modic, K. A. and Nicklas, Michael and König, Markus and Low, D. and Ghosh, Sayak and Mackenzie, Andrew P. and Arnold, Frank and Hassinger, Elena and McDonald, Ross D. and Winter, Laurel E. and Bauer, Eric D. and Ronning, Filip and Ramshaw, B. J. and Nowack, Katja C. and Moll, Philip J. W.},
  journal  = {Science},
  title    = {Spatial control of heavy-fermion superconductivity in {CeIrIn}$_{5}$},
  year     = {2019},
  issn     = {0036-8075, 1095-9203},
  month    = oct,
 number_shit  = {6462},
  pages    = {221--226},
  volume   = {366},
  doi_shit        = {10.1126/science.aao6640},
  language = {en},
  shit_url      = {https://www.sciencemag.org/lookup/doi/10.1126/science.aao6640},
  shit_urldate  = {2021-06-15},
}

@Article{Carneiro2000,
  author     = {Carneiro, Gilson and Brandt, Ernst Helmut},
  journal    = {Phys. Rev. B},
  title      = {Vortex lines in films: {Fields} and interactions},
  year       = {2000},
  month      = mar,
  note_shit       = {Publisher: American Physical Society},
 number_shit    = {9},
  pages      = {6370--6376},
  volume     = {61},
  doi_shit        = {10.1103/PhysRevB.61.6370},
  shorttitle = {Vortex lines in films},
  shit_url        = {https://link.aps.org/doi/10.1103/PhysRevB.61.6370},
  shit_urldate    = {2020-12-02},
}

@Book{Poole2007,
  author    = {Poole, C.P. and Farach, H.A. and Creswick, R.J. and Prozorov, R.},
  publisher = {Elsevier},
  title     = {Superconductivity},
  year      = {2007},
}

@Article{Machida2012,
  author   = {Machida, Y. and Itoh, A. and So, Y. and Izawa, K. and Haga, Y. and Yamamoto, E. and Kimura, N. and Onuki, Y. and Tsutsumi, Y. and Machida, K.},
  journal  = {Phys. Rev. Lett.},
  title    = {Twofold {Spontaneous} {Symmetry} {Breaking} in the {Heavy}-{Fermion} {Superconductor} {UPt}$_{3}$},
  year     = {2012},
  issn     = {0031-9007},
  month    = apr,
  note_shit     = {Num Pages: 5 Place: College Pk Publisher: Amer Physical Soc Web of Science ID: WOS:000302635600012},
  number_shit   = {15},
  pages    = {157002},
  volume   = {108},
  doi_shit      = {10.1103/PhysRevLett.108.157002},
  language = {English},
  url_shit      = {https://www.webofscience.com/api/gateway?GWVersion=2&SrcAuth=PQPLP&SrcApp=WOS&DestURL=https%3A%2F%2Fwww.proquest.com%2Fdocview%2F2082257658%2Fembedded%2FSEY249C5IAC2N1J8%3Fpq-origsite%3Dwos&DestApp=PQP_ExternalLink&SrcItemId=WOS:000302635600012&SrcAppSID=EUW1ED0C3FK4kbCaoakT1FmWjjgTP},
  urldate  = {2024-03-25},
}

@Article{Hasselbach1989,
  author   = {Hasselbach, K. and Taillefer, L. and Flouquet, J.},
  journal  = {Phys. Rev. Lett.},
  title    = {Critical point in the superconducting phase diagram of UPt$_{3}$},
  year     = {1989},
  month    = jul,
  note_shit     = {Number: 1},
  number_shit   = {1},
  pages    = {93--96},
  volume   = {63},
  doi_shit      = {10.1103/PhysRevLett.63.93},
  url_shit      = {https://link.aps.org/doi/10.1103/PhysRevLett.63.93},
  urldate  = {2020-01-07},
}

@Article{Huang2014,
  author   = {Huang, Wen and Taylor, Edward and Kallin, Catherine},
  journal  = {Phys. Rev. B},
  title    = {Vanishing edge currents in non- p -wave topological chiral superconductors},
  year     = {2014},
  issn     = {1098-0121, 1550-235X},
  month    = dec,
  number_shit   = {22},
  pages    = {224519},
  volume   = {90},
  doi_shit      = {10.1103/PhysRevB.90.224519},
  language = {en},
  url_shit      = {https://link.aps.org/doi/10.1103/PhysRevB.90.224519},
  urldate_shit  = {2021-03-08},
}

@book{Mineev_introduction_1999,
	title = {Introduction to {Unconventional} {Superconductivity}},
	isbn = {90-5699-209-0},
	publisher = {Gordon and Breach Science Publishers},
	author = {Mineev, V.P. and Samokhin, K.V.},
	year = {1999},
}

@article{wuchner_magnetic_1993,
	title = {Magnetic properties of the heavy-fermion superconductors {UPt$_{3}$} and {URu$_{2}$Si$_{2}$}},
	volume = {85},
	issn_shit = {0038-1098},
	url_shit = {https://www.sciencedirect.com/science/article/pii/003810989390032I},
	doi_shit = {10.1016/0038-1098(93)90032-I},
	language = {en},
	number = {4},
	urldate_shit = {2021-07-27},
	journal = {Solid State Communications},
	author = {Wüchner, S. and Keller, N. and Tholence, J. L. and Flouquet, J.},
	month = jan,
	year = {1993},
	pages = {355--360},
	file_shit = {ScienceDirect Full Text PDF:/home/hasselbach-ubuntu/snap/zotero-snap/common/Zotero/storage/APHYWHFY/Wüchner et al. - 1993 - Magnetic properties of the heavy-fermion supercond.pdf:application/pdf;ScienceDirect Snapshot:/home/hasselbach-ubuntu/snap/zotero-snap/common/Zotero/storage/RQKUI3WZ/003810989390032I.html:text/html},
}
\bibliographystyle{sciencemag}
\renewcommand\thefigure{{S\arabic{figure}}}    
\setcounter{figure}{0} 

\section*{Acknowledgments}
The authors thank J.P. Brison, J.R. Kirtley, M. K. Arfaoui and T. Winyard for useful discussions and T. Crozes for the SQUID fabrication.
\paragraph*{Funding:}
This work is supported by the French National Research Agency in the framework of the investissments d’avenir program (ANR-15-IDEX-02)
 ANR-22-CE30-0040-02, PGC was funded by the Nanosciences Foundation in Grenoble (France) and the GreQuE Cofund programme 810504
\paragraph*{Author contributions:}
P.G.C. carried out the SSM experiment, analyzed the data and contributed to the writing, V.O.D. carried out the first SSM experiments and contributed to the writing, M-K. A. carried out the most recent SSM experiments and contributed to the writing A.D.H. and D.A. grew and characterized the crystals and contributed to the writing,
Scanning SQUID Data was analyzed by P.G.C., V.O.D., M.-K. A., A.D.H and K.H. The project was  supervised by K.H. and he contributed to the writing.
 
\paragraph*{Competing interests:}
There are no competing interests to declare.
\paragraph*{Data and materials availability:}
CNRS openData serveur


\subsection*{Supplementary materials}
Materials and Methods\\
Supplementary Text\\
Figs. S1 to S10\\
References \textit{(7-\arabic{enumiv})}\\ 


\newpage


\renewcommand{\thefigure}{S\arabic{figure}}
\renewcommand{\thetable}{S\arabic{table}}
\renewcommand{\theequation}{S\arabic{equation}}
\renewcommand{\thepage}{S\arabic{page}}
\setcounter{figure}{0}
\setcounter{table}{0}
\setcounter{equation}{0}
\setcounter{page}{1} 


\begin{center}
	\section*{Supplementary Materials for\\ \scititle}
P. Garc\'ia Campos$^{1,\dagger}$,
V.O. Dolocan$^{1,3,\dagger}$,
M.-K. Arfaoui$^{1,2}$, 
D. Aoki,$^{5,\dagger}$,
A. D. Huxley$^{4,\dagger}$,\\
K. Hasselbach$^{1,\dagger,\ast}$\\
\normalsize{$^\ast$To whom correspondence should be addressed; E-mail: klaus.hasselbach@neel.cnrs.fr}

	\small$^\dagger$These authors contributed equally to this work.
\end{center}

\subsection*{Materials and Methods}
\subsubsection*{Scanning SQUID Microscopy setup}
The microscope operates in a reversed dilution refrigerator. The micrometer sized SQUID is scanned over the sample surface.The SQUID chip is attached to a home built scanning force microscope with a quartz tuning fork as force detector. The difference in the length of the DC-SQUID branches makes the critical current-flux transfer function unambiguous at zero flux.
A room temperature copper solenoid is used to apply magnetic field. Acquiring an image takes between 15 and 30 minutes depending on the image size. The fast scanning direction is the y direction for the Figures 1-5, S6 and S7 fast scanning is along the x direction.

\begin{figure}[H]
	\centering
	\includegraphics[width=0.7\linewidth]{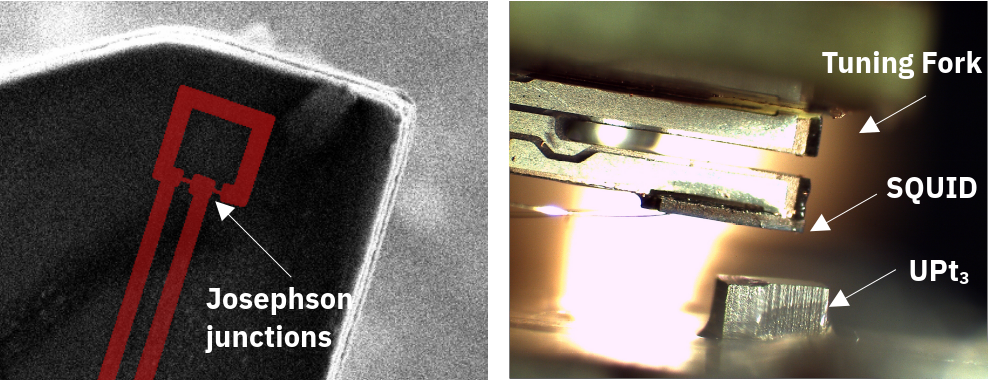}

	\caption{\textbf{SQUID-Probe and microscope setup}. Left: Aluminum SQUID of 1$\mu$m diameter at 1$\mu$m distance from an edge of a Si chip. Right the Si chip fixed on a quartz tuning fork is ready to scan the surface of a \UPt single crystal.}
	\label{fig:s1}
\end{figure}

\subsubsection*{Crystal}

A single crystal ingot of \UPt was grown by Czochralski Method (RRR=580). A parallelepiped  (1.5x1x0.65 mm$^{3}$) was spark cut from the ingot, polished and annealed at \SI{950}{\degreeCelsius} for seven days in UHV. The specific heat of the sample was measured using a PPMS QD \Hethree system with T$_{c+}$ 
of 0.514K and transition width 25 mK (Fig. S4).
The transition from the A- to the B-phase occurs at 0.46K with width $\sim$ 25 mK.

\subsubsection*{Monopole model of the magnetic stray field of a vortex}
In a conventional type-II superconductor, the magnetic field induces vortices with one  $\Phi_{0}$.
The magnetic field profile of the vortex's z-component of the stray field above a superconductor can be expressed by the stray field of a magnetic monopole:
\begin {equation}\label{eq:1}
B_{z}(r,T) = \frac{\Phi_{0}}{2\pi}
\frac{(z + 1.27\lambda(T))}{(r^{2} + (z + 1.27\lambda(T))^{2} )^{3/2}},
\end {equation}
z being the height above the sample surface, $\lambda$ the magnetic penetration depth and r the radial distance from the vortex center. The factor 1.27 has been determined \cite{Carneiro2000} in order to describe the magnetic field close to the vortex.

 In \ref{s_cut_x_cut_y}(A,B) we show two orthogonal magnetic field profiles of one vortex acquired at 0.3K. The field profiles coincide, demonstrating that any deviation from cylindrical symmetry of the field distribution is too small to be detected in our measurements.  It is impossible to determine independently the penetration depth and SQUID sample distance (height) from the formula EQN \ref{eq:1}. The fitting parameters are a field offset (not included in EQN \ref{eq:1}) and 1.27$\lambda+z$. 
 The offset is 5 $\mu$T and 1.27$\lambda$+z=3.02$\mu$m. For half-quantized vortices $\phi_0$ is replaced with $\phi_0$ /2.

 \begin{figure}[H]
 	\centering
 	\includegraphics[width=0.9\linewidth]{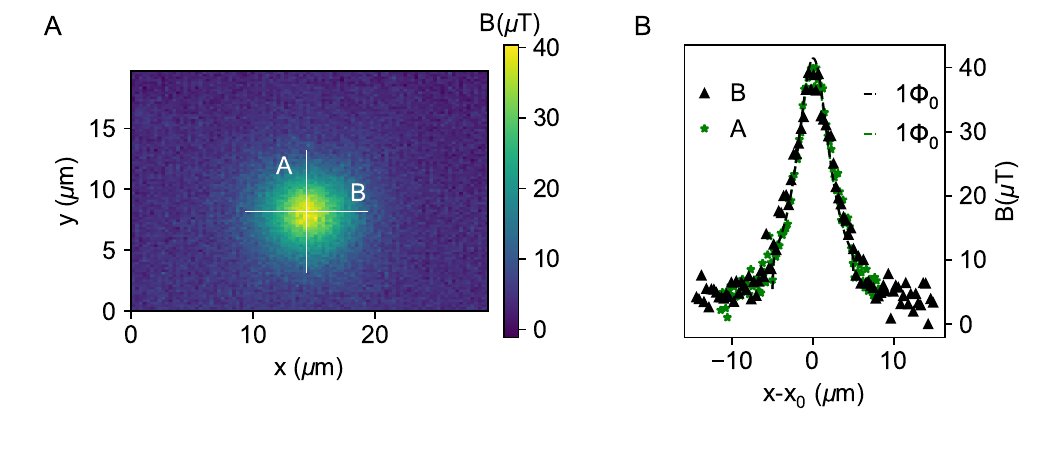
  }
 	\caption{\textbf{Characterization of a single vortex \UPt}. Panel (A) is the magnetic image of a single vortex in close to zero applied magnetic field. In panel (B) flux profiles along the axes labeled A and B in panel (A) are shown. The dash dotted lines are fits to the monopole model.}
 	\label{s_cut_x_cut_y}
 \end{figure}

\subsubsection*{Sample properties: resistivity, specific heat, susceptibility} 
\begin{figure}[H]
	\centering
	\includegraphics[width=0.7\linewidth]{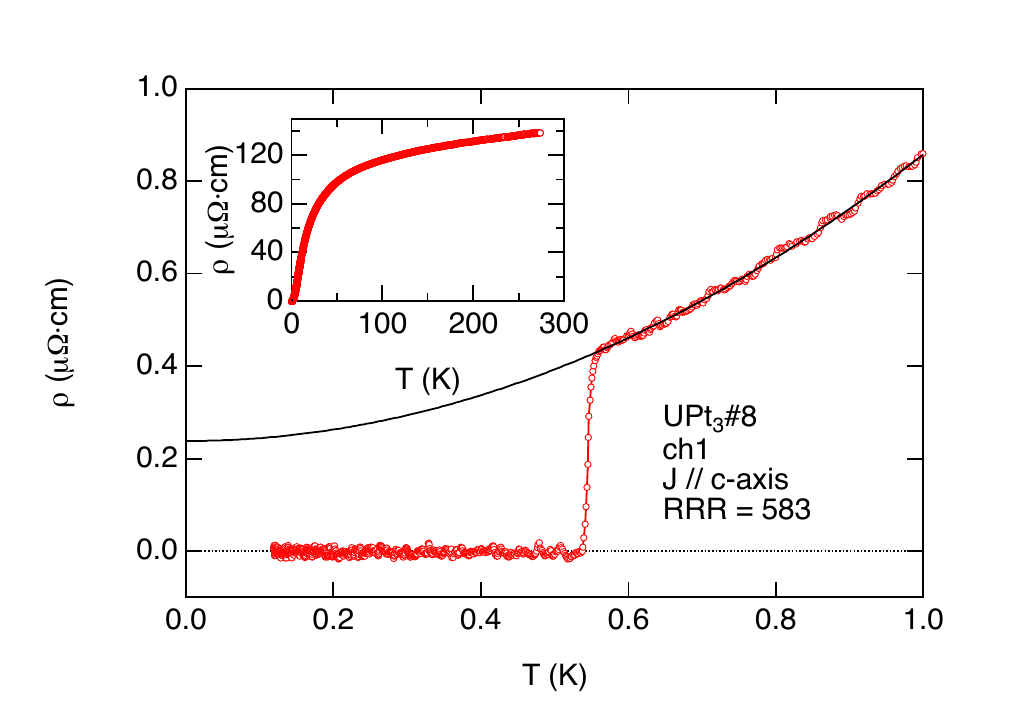}
	\caption{ \textbf{Temperature dependence of electrical resistivity of \UPt crystal}. Resistivity of a \UPt sample cut from the same Ingot as the crystal studied, measured at the CEA Grenoble. The onset of the superconducting transition is at 0.556K and the resistance is zero below 0.54K.}
	\label{s_Tc_ro_Dai}
\end{figure}

\begin{figure}
	\centering
	\includegraphics[width=0.7\linewidth]{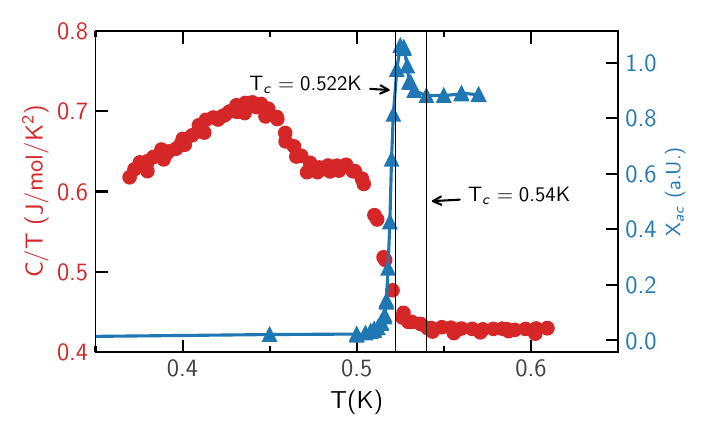}
	\caption{\textbf{Determination of the transition temperatures of the A and B phases for the given \UPt crystal}.Specific heat of the \UPt crystal used in the SSM measurements at zero magnetic field, measured in a ${}^{3}$He PPMS system from Quantum Design. The blue curve is the local susceptibility of the crystal as a function of temperature. The later measurement was obtained with the scanning SQUID microscope setup. A room temperature solenoid applied magnetic fields of $\pm$ 50 $\mu$T. The difference in the SQUID signal for $+$  and $-$ 50 $\mu$T applied field corresponds to a local susceptibility.}
	\label{s_Tc_cp_xac}
\end{figure}

\begin{figure}[!t]
	\centering
    \includegraphics[width=0.7\linewidth]{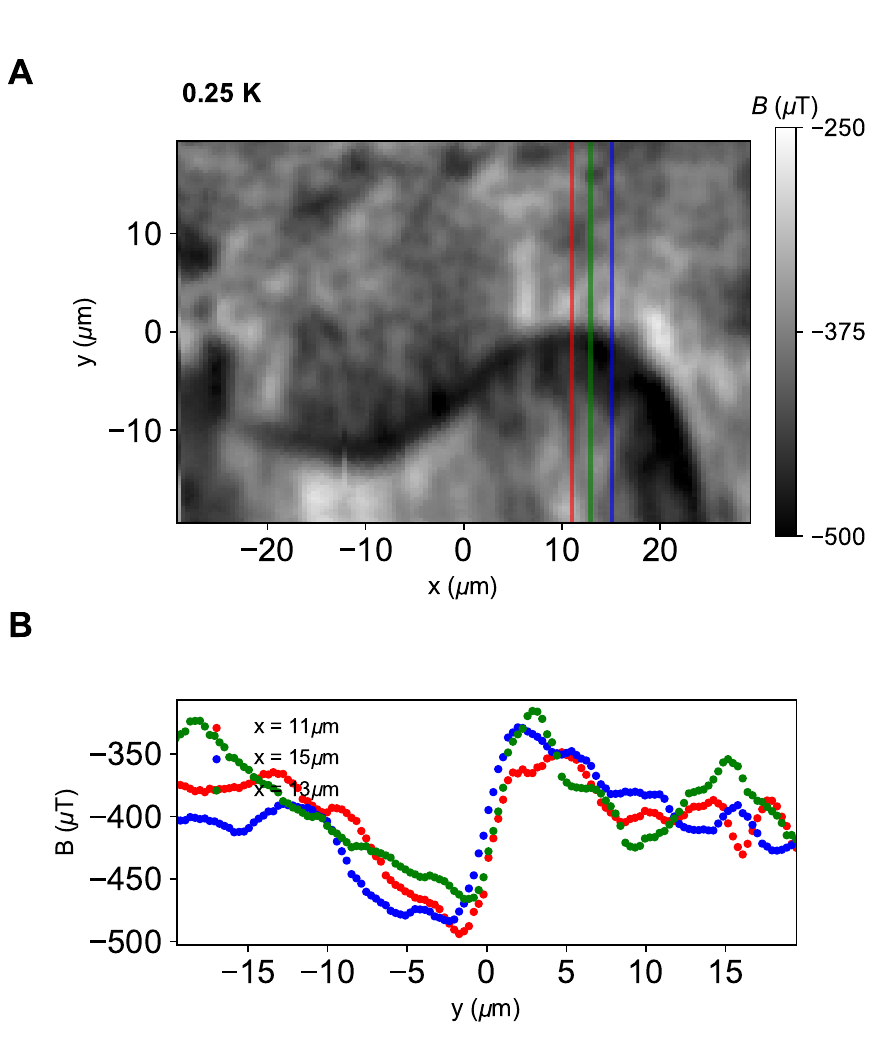}
	\caption{\label{S6_250mK} \textbf{Visualization of vortices and vortex sheet}
	Fig A: Larger image of Fig.\ref{Fig_4}(D) (T=0.25 K and H= -0.4\,mT/$\mu_{0}$. Vortices appear as dark spots. The domain wall is a region of highly in-homogeneous and intense magnetization.\\
	Fig (B) 3 line profiles across the domain wall, the variation in magnetization at the domain wall is between 0.1- 0.15\,mT.
 The average magnetization is close to the magnetic field (-0.4\,mT/$\mu_{0}$) applied prior to cooling below T$_{c}$.
 Weak flux expulsion of 2-3\% \cite{wuchner_magnetic_1993,Vincent1991} has been observed in UPt$_{3}$.
 }
	\label{s_zoom_vortex_vortex_sheet}
\end{figure}

\clearpage
\subsubsection*{Other crystal}
Scanning SQUID images acquired several years earlier and shown in Figs. S6 and S7 taken on a different crystal (its heat capacity is shown in Fig. S8) \cite{Dolocan2005}. 
The microscope used was the first generation microscope \cite{Dolocan2005}. The SQUID had an inner dimension of 1$\mu$m, the lead inductances were balanced, at zero flux such that the SQUID's critical current was maximum. Thus transforming the periodic critical current modulation to magnetic induction becomes ambiguous, for example in Fig \ref{fig:s_pos_neg_Dolocan}(C) at the lower left corner shows a minimum in the yellow region at the dark circular domain, this is attributed to the lowering of the critical current away from its maximum at zero flux. Horizontal streaks in the images are due to the SQUID touching the crystal surface.
At the time the microscope was not equipped with coarse displacements in the x, y plane.  Images of the same cool down were all acquired within the range of the sample scanner.

The images confirm that the formation of vortex structures is a generic property of the superconducting state in the B-phase of \UPt single crystals. These images have been taken subsequently during the same cool down.

\subsubsection*{Domain wall energy; experimental estimate}

In magnetic field $B/\mu_0$ the vortex spacing in the bulk (for a hexagonal lattice) $a$ is
\begin{align}
a= \sqrt{\frac{2 \phi_0}{\sqrt{3} B}}. 
\end{align}
For a magnetic field of 450 $\mu$T/$\mu_0$, $a=2.3 \mu\text{m}$.

To estimate the magnetic pressure exerted on the domain wall from our measurements, the distance over which the magnetic field changes across the wall is needed. This distance is too short to be resolved in our measurements. The penetration depth $\lambda \sim 0.75 \mu \text{m}$ at $0.25~\text{K}$ (see later) implying that the field gradient at the wall is $dB/dz \sim \Delta B/\lambda$ with $\Delta B$ the step in the field either side of the wall. This induces a current density $J=\frac{\Delta B}{\mu_0 \lambda}$. The magnetic pressure on the wall is calculated from the flux per unit length $\Gamma \phi_0/a$ multiplied by $J$ 
\begin{align}
P=\frac{1}{\mu_0} \frac{dB}{dz} \frac{\phi_0}{a} \Gamma
\end{align}
with $\Gamma$ the enhancement of the flux density along the wall compared with in the bulk. Some of our measurements indicate a slight excess of flux at the domain wall, but this is not more than 20\% above the bulk value, suggesting $\Gamma \sim 1.2$.  

The measured field change across the wall is $\Delta B_\text{mes} \sim 150 \mu\text{T}$ (Fig S5(B)). For a surface-SQUID distance of $1.8 ~\mu \text{m}$ the change at the surface is approximately double $\Delta B_\text{mes}$ (see below and Fig. S9A).  For $\Delta B \sim 300 \mu\text{T}$ and a radius of curvature $R \sim 14~ \mu\text{m}$ from Fig. S5(B), with $\Gamma=1.2$, $P=0.34~ \text{Nm}^{-2}$ and $\gamma_\text{expt} = P R = 5 \times 10^{-6} ~\text{Jm}^{-2}$.

\subsubsection*{Domain wall energy; theory estimate}

In the Ginzburg-Landau expansion, the homogeneous part of the free energy density for a two component order parameter $\bm{\eta}=\left\{\eta_1,\eta_2\right\}$ appropriate to \UPt is
\begin{align}
\mathcal{F} &= -\alpha \bm{\eta} \cdot \bar{\bm{\eta}} + \beta_1 (\bar{\bm{\eta}}\cdot \bm{\eta})^2 + \beta_2 |\bm{\eta} \cdot \bm{\eta}|^2
&= \frac{\alpha^2}{\beta_1}\left( - \frac{1}{2}(|\psi_+|^2 + |\psi_-|^2) + \frac{1}{4}(|\psi_+|^4 + |\psi_-|^4) + \beta_\perp |\psi_+|^2 |\psi_-|^2 \right)
\label{eq:F}
\end{align}
with 
$
\psi_\pm =\sqrt{\frac{\beta_1}{\alpha}}\left( \eta_1 \pm i\eta_2 \right),
$
and 
$
\beta_\perp= \frac{1}{2}+\frac{\beta_2}{\beta_1}$. $\bar{\bm{\eta}}$ is the complex conjugate of $\bm{\eta}$ and the dot product denotes a sum over products of the same component.

This results in a TRS broken chiral state for $\beta_\perp>\frac{1}{2}$, $\beta_2/\beta_1>0$.  The SBF is not included in the above expression. It results in two transitions from which $\beta_\perp$ can be determined from \cite{Joynt2002}
\begin{align}
\frac{\beta_2}{\beta_1}\approx \frac{(\Delta C^-/T_c^-)}{(\Delta C^+/{T_c^+})}
\end{align}
with $\Delta C^\pm$ the jumps in heat capacity at $T_c^\pm$.
For the heat capacity data in Fig. S4, $\beta_2/\beta_1=0.54(2)$.

The uniform Free energy density is minimized for a state $|\psi_+|=0, |\psi_-|=1$ or $|\psi_+|=1, |\psi_-|=0$ with $\mathcal{F}  \equiv \mathcal{F}_0=-\alpha^2/4\beta_1$. A domain wall switches between these two minima. At the domain wall $|\psi_+|=|\psi_-|$. With this constraint the minimum condensation energy at the wall is increased by 

\begin{align}
\frac{\Delta\mathcal{F}}{\mathcal{F}_0}= \frac{\beta_2/\beta_1}{1+\beta_2/\beta_1}.  
\end{align}
Since $\kappa$ is large ($\kappa \sim 50$ for \UPt) the potential energy density at the center of the wall can be determined independently of the gradient energy.  
If there are no gradient coupling terms the gradient energy makes an identical contribution to the condensation energy. The wall energy is then
\begin{align}
\gamma_\text{theory}= \frac{2~ t~ \mathcal{F}_0 ~\beta_2/\beta_1}{1+\beta_2/\beta_1}  
\label{S7}
\end{align}
with $t$ the length scale over which the order parameter is suppressed.  
The above formula is used in the main text with $\mathcal{F}_0$ replaced with the BCS estimate of the uniform state condensation energy.  
If gradient mixing terms are present or for an applied magnetic field there will be additional contributions. 
For Sauls'\cite{Sauls1994} $E_{2u}$ model the gradient coupling parameters are estimated to be close to zero. 

The order parameter changes on the length scale $t\sim \xi \sim 110\text{\AA}$ with $\xi$ the coherence length, which also determines the radius of the vortex cores. The energy gain from locating vortices on the wall is then $(2 \xi/a) t \Delta \mathcal{F}$, which can be neglected compared with the preceding estimate of the wall energy. 

The expression \ref{S7} evaluates to $\gamma_\text{theory}= 5\times10^{-6}~\text{Jm}^{-2}$.

\subsubsection*{Estimate of the field from chiral currents}

Chiral currents arise from gradient coupling terms in the Ginzburg-Landau theory, resulting in a sharp step (on the length scale of $\sim \xi$) in the magnetic field across a wall. 
The Gradient terms in the Free energy are
\begin{align}
\mathcal{F}_\text{grad}\Large{/}\frac{\alpha K_1}{\beta_1} &=  \sum_\pm \overline{\psi_\pm} \left(-\frac{1+C_1}{2}\left( D_x^2+D_y^2\right) \pm C_2 B \right) \psi_\pm
- \frac{C_1}{2}\sum_\pm \overline{\psi_\pm}\left(D_x \pm i D_y\right)^2 \psi_\mp
\end{align}
With gradient coupling coefficients $C_1 = (K_2+K_3)/ (2 K_1)$ and $C_2 = (K_2-K_3)/(2 K_1)$ where $K_1,K_2,K_3$ are the parameters defined in the literature \cite{Sauls1994,Joynt2002}. $D = \nabla +i (2 \pi/\phi_0)A$ and the bar denotes a complex conjugate. Different models give different coupling values. Most have $C_2 \sim 0$, but differ in the assignment of $C_1$ with $C_1 \sim 0$ for Sauls' $E_{2u}$ model while $C_1 \sim 1$ for the $E_{1g}$ model ($C_1>1$ would result in spontaneously modulated state and require retaining terms beyond quadratic in the gradients). $C_2 \ne 0$ couples the order parameter directly to the magnetic field and would result in the superconductivity having an intrinsic moment. We take $C_2$ to be zero in the following. $C_1$ couples the field to the gradient of the order parameter \cite{Benfenati2020}. Assuming translational invariance parallel to the domain wall (perpendicular to $x$) the order parameter can be parameterized by $(\psi_+,\psi_-)=\rho(x)\left(\cos[\theta(x)], \sin[\theta(x)]e^{i \phi}\right)$ where $\theta$ changes from $\theta=0$ at $x=-\infty$ to $\theta=\pi/2$ at $x=+\infty$. $\rho(x)$ is symmetric about the domain wall with zero derivative at $x=0$, whereas $\theta(x)-\theta(0)$ is anti-symmetric. In the large $\kappa=\lambda \sqrt{|\alpha/K_1|}$ limit $\rho$ is then determined by minimizing the homogeneous free energy at each position and can be replaced by $\rho(\theta)$. The Free energy plus magnetic energy $B^2/2\mu_0$ is then minimized with respect to the vector potential $A$ (and $\phi$). Integration of the resulting expression gives the field step across the domain wall
\begin{align}
\Delta B = \frac{\pi}{2} \frac{C_1}{\sqrt{1+ \frac{\beta_2}{\beta_1}}}  \frac{\phi_0}{2 \pi \lambda^2} 
\end{align}
with
\begin{align}
\lambda = \frac{\phi_0}{2\pi} \sqrt{\left|\frac{\beta_1}{\mu_0 \alpha K_1}\right|}.
\end{align}
The field profile is screened over a length $\lambda_\text{p} \sim \lambda/\sqrt{1+C_1}$ far from the wall (when the order parameter approaches its uniform value). For $C_1\ne 0$ magnetic field variations at shorter distances are no longer characterized by a single length scale. Nevertheless, experimental results are often interpreted in such terms. Values can be deduced from the magnetic contrast observed with small-angle neutron scattering. This was measured for $B \sim 0.2 \text{T}$, which should be compared with the the $B\rightarrow C$ transition field $B_\text{BC}/\mu_0> 1~\text{T}$ for $T=0-400~\text{mK}$ \cite{Huxley2000,Gannon2015}. Assuming a constant value for the coherence length $\xi \sim 110  \text{\AA}$ gives $\lambda_p \sim 10,000 ~\text{\AA}$ at 350 mK. A slightly shorter value of $\lambda_p \sim 9000 ~\text{\AA}$ was obtained at the same temperature and 18 mT with $\mu SR$ based on the standard deviation of the field distribution (the same study gives $\lambda_p\sim 7500 \text{\AA}$ at 250 mK, used in the previous section)  \cite{Yaouanc1998}.    

If we take $C_1 \sim 1$ and $\lambda_\text{p} \sim 9000~\text{\AA}$, $\Delta B \sim 250 ~\mu\text{T}$.  This is the value in the bulk and might be expected to be strongly screened approaching the sample surface. Chiral currents are also not topologically protected and are expected to be suppressed or strongly reduced in f-wave superconductors (considered to be the dominant orbital pairing channel in UPt$_3$) compared to p-wave ones\cite{Huang2014,Kallin2016}.

\subsection*{Measured field profiles}

In our geometry the surface is perpendicular to the c-axis and the applied field $H_\text{app} \parallel c$. The field deep in the bulk is thus parallel to $c$. An accurate calculation of how the magnetic field at the surface is related to that deep inside the bulk requires a detailed model for screening inside the sample similar to the analysis for a single vortex in a conventional superconductor \cite{Carneiro2000}. The field at height $z$ can then be found from the surface field by a straightforward application of magneto-statics.

In this section we consider two approximations to relate the field deep in the bulk to that at the surface. We assume that either (i) the current density is independent of the distance $z$ from the surface inside the superconductor or (ii) the field $B_z$ in the direction normal to the surface is independent of $z$ inside the superconductor i.e. $B_z(x,z=0)=B_{z}(x,\text{bulk})$. For approximation (i) the ratio $B_z(x,0)/B_{z}(x,\text{bulk})=1/2$. The resulting profile at arbitrary height above the surface $z$ relative to the field at the surface is calculated from a standard application of magneto-statics,
\begin{align}
B_z(x,z)=\frac{1}{\pi} \int_{-\infty}^\infty dx'~\frac{B_z(x',0)~z}{(x-x')^2+z^2}.    
\end{align}

Approximation (ii) is expected to better describe the field profile from a dense array of flux lines since all the magnetic flux has to exit the sample across the surface. 
The resulting profile for the field at height $z$ from the surface is shown in Fig. S9(A), with parameters appropriate to the scans in Fig S5(B).  This shows that the peak field at the surface (equal to that in the bulk for this approximation) is approximately double the peak field measured by the SQUID at $z=1.8 \mu\text{m}$ in Fig. S5(B).


The field due to chiral currents does not carry a net flux. Approximation (i) (with constant currents) is therefore more appropriate.  Setting the background field to 0,
the field step from $-\Delta B/2$ to $+\Delta B/2$ across the wall occurs over the coherence length (much too short to resolve in our measurements) with the field either side of wall screened over the length scale $\lambda_p$. For the SQUID distance in Fig. S5(B) the field step detected due to chiral currents is therefore expected to be reduced from the step at the sample surface by a factor of over $16$ (Fig S9B). The large reduction is due to the smaller length scale of the field variation. Assuming the chiral currents retain their bulk profile up to the surface, the value of the field change detected by the SQUID relative to the bulk field profile is reduced by a further factor of 2.  The detected field step would then be $8 \mu\text{T}$ across the wall for $C_1=1$. In practice the screening may be stronger than assumed in this estimate, resulting in a smaller field step. 

\begin{figure}[!t]
    \centering
	\includegraphics[width=0.5\linewidth]{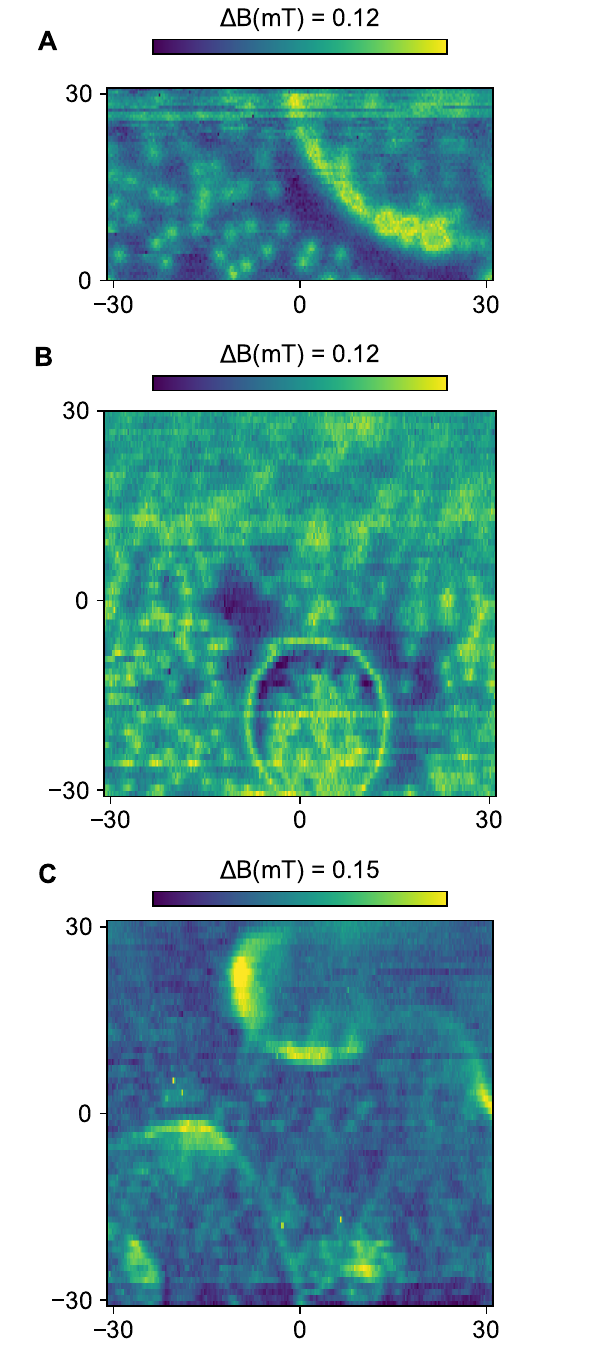}\\
	\caption{  \textbf{Coexistence of vortices and domains in crystal B.} Formation of circular magnetic flux structures in UPt$_{3}$: (A) Magnetic image taken after applying 0.04\,mT/$\mu_{0}$ FC at a T$\approx$0.35K. The dark blue region close to the circular domain wall is void of vortices. The dimensions of the image are 62 $\times 31 \mu$m. (B) Subsequent scan after field cooling under 0.1\,mT/$\mu_{0}$. The dimensions of the image are 62  $\mu$m $\times$ 62 $\mu$m. (C) Subsequent scan after field cooling under 0.2\,mT/$\mu_{0}$, at T$\approx$0.35K. The dimensions of the image are 62$\mu$m $\times$ 62$\mu$m}.
 \label{fig:s_pos_Dolocan}
  \end{figure}

\begin{figure}[!t]
 \centering
	\includegraphics[width=14cm]{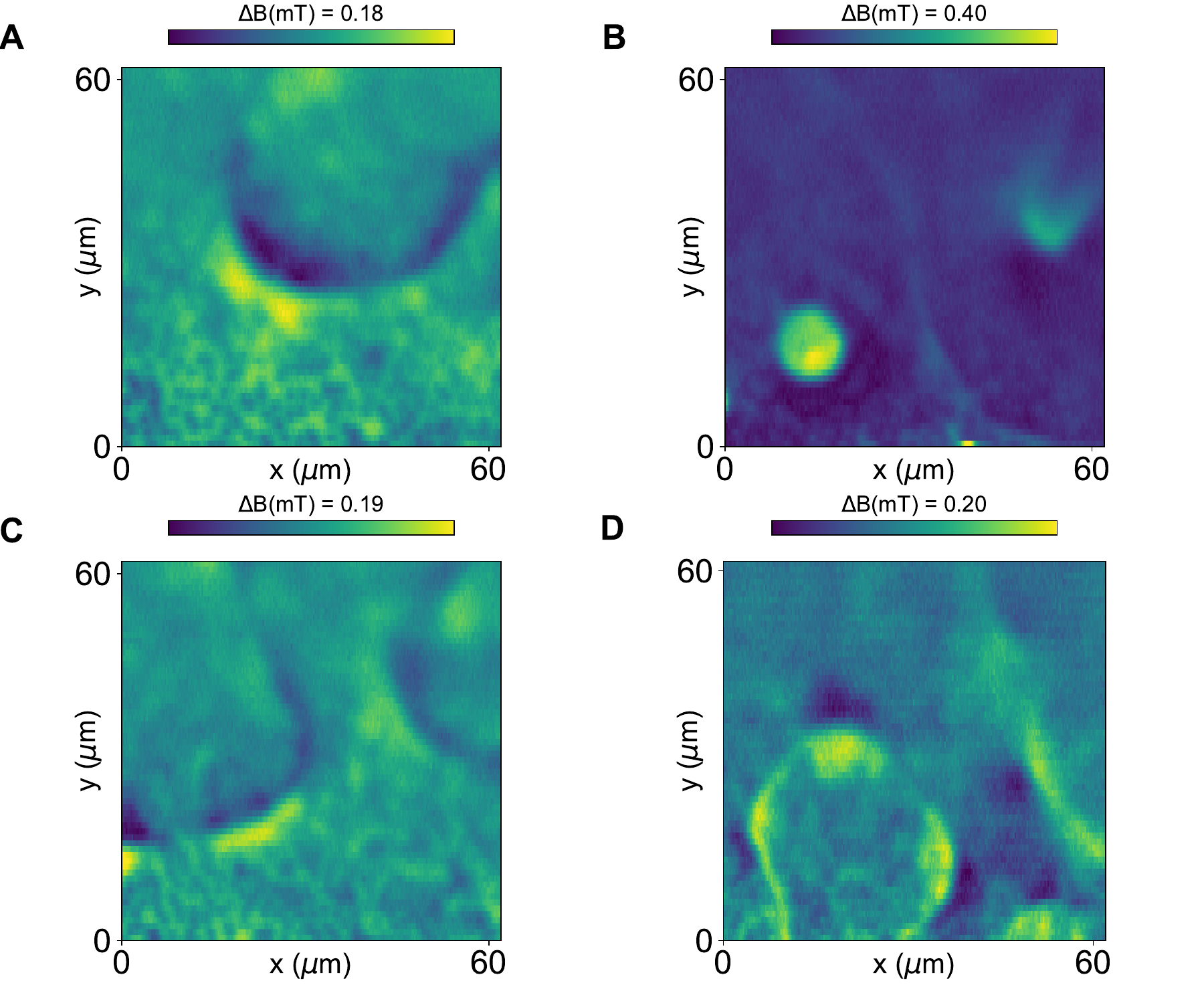}\\
	\caption{ \textbf{Vortex sheet and domains depend on direction and amplitude of the magnetic field during field cooling, crystal B} Formation of domain walls for different applied fields during field cooling into the B-phase, T$\approx$0.35K: (A) -0.3\,mT/$\mu_{0}$, (B) 0.3\,mT/$\mu_{0}$, (C) -0.4\,mT/$\mu_{0}$, (D) 0.4\,mT/$\mu_{0}$. Images (A-D) are from the same cool-down at the same location.}
 \label{fig:s_pos_neg_Dolocan} 
\end{figure}

\begin{figure}[!t]
\centering
	\includegraphics[width=9cm]{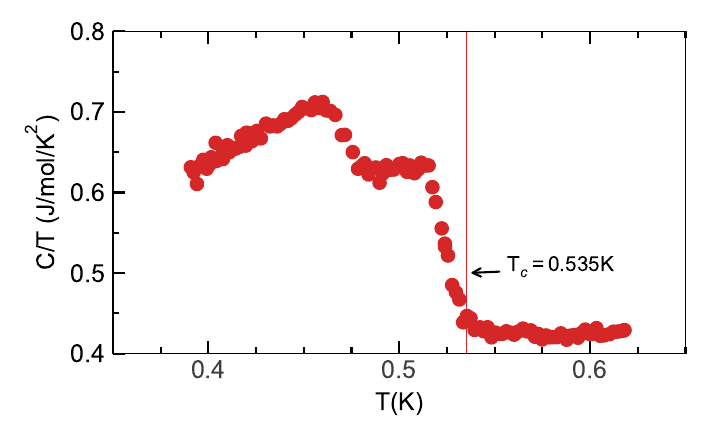}\\
	\caption{ \textbf{Specific heat with double transition of crystal B.} Specific heat presented as C over T as function of T for the sample presented in the supplementary data, acquired using ${}^{3}$He PPMS system from Quantum Design at CEA Grenoble.}
  \label{fig:s_cp_huxley}
\end{figure}

\begin{figure}[!t]
\centering
	\includegraphics[width=16cm]{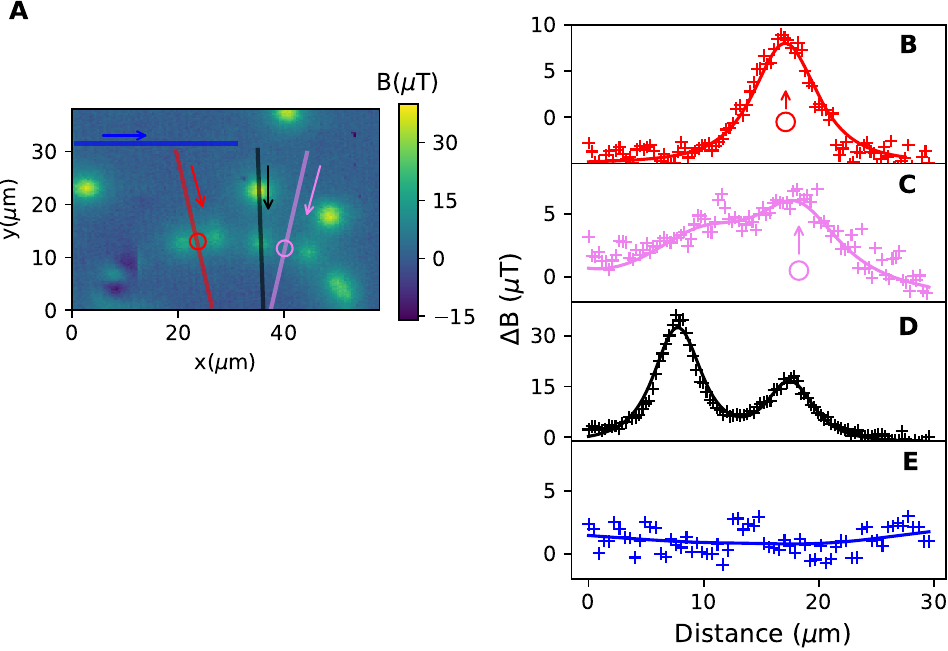}\\
 \caption{\textbf{Chiral currents are not observed.} Four line profiles are presented in (A), same image as Fig \ref{Fig_3} (B) (T=0.35 K), the arrows indicate the sens. The red, pink and black lines pass through the half vortex alignment. The blue line is a reference line.
 Crosses are data points. (B,C) the red and pink circles indicate the position of the half vortex alignment. The red and pink lines are calculated profiles using the monopole model with a penetration depth + SQUID sample distance of 3.04 $\mu$m and offset of -5 $\mu$T (B) an -2  $\mu$T (C). (D) The black line is a fit of two vortices 1$\Phi_{0}$ and 0.5$\Phi_{0}$ with a penetration depth + SQUID sample distance of 3.04 $\mu$m. (E) The blue crosses represent a reference line, indicating a noise in the scans  of the order of $\pm$ 1.5$\mu$T. In the limit of the resolution the line profiles can be modeled with a superposition of the stray field of vortices.
 }
 \label{fig:S11}
 \end{figure}

\begin{figure}[!t]
\centering
	\includegraphics[width=16cm]{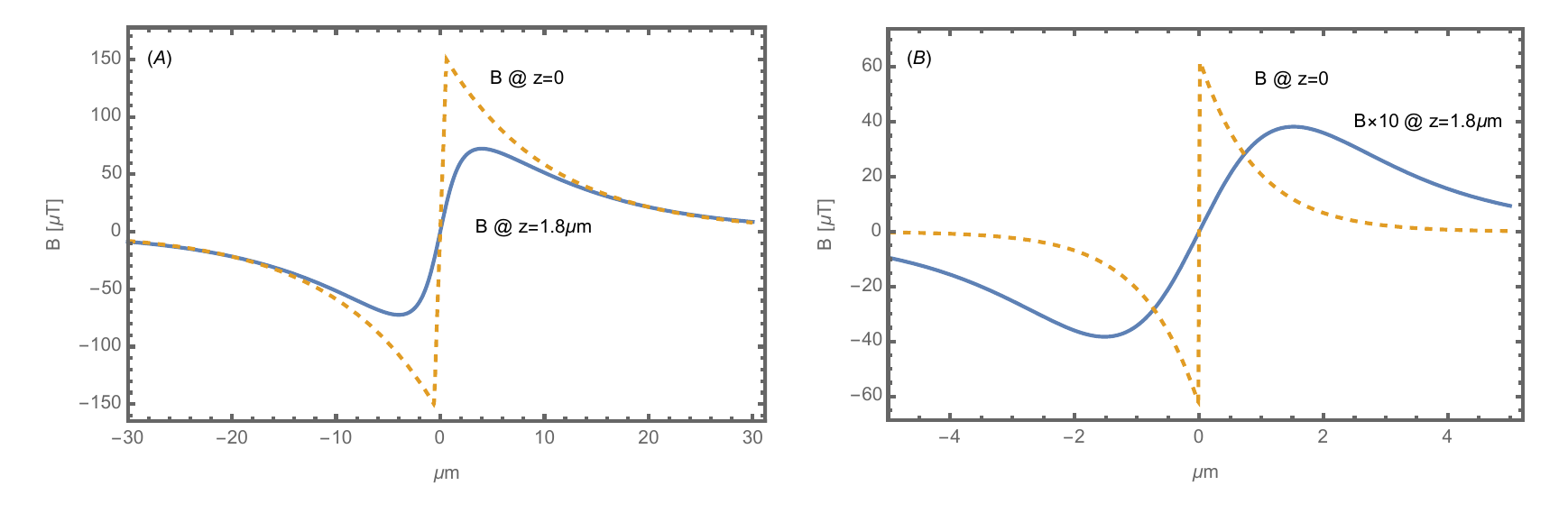}\\
	\caption{Shows magnetic profiles normal to the surface at two different distances from the surface.  The dashed curves show a profile at $z=0$, the solid curves give the corresponding calculated field at the SQUID height $z=1.8 ~\mu \text{m}$. In \textbf{Panel (A)} the shape of the profile at $z=1.8 ~\mu \text{m}$ is similar to that in Fig S4B, showing that the measured peak field is approximately a factor of 2 smaller than the field at the sample surface. \textbf{Panel (B)} shows the result for a much sharper field profile of the type predicted due to chiral currents screened over a distance $\lambda_p=0.9 \mu\text{m}$. The field profile at $z=1.8~\mu \text{m}$ has been multiplied by 10. The amplitude of the field step at the SQUID position is reduced by a factor of just over 16 relative to that at the surface in this case.}
  \label{fig:S9}
\end{figure}
\end{document}